\documentclass[twocolumn,twocolappendix]{aastex63}

\usepackage{amsmath}

\usepackage{pifont}
\submitjournal{ApJ}

\shorttitle{CKD for Escape Sim.}

\begin{document}

\title{Numerical performance of correlated-k distribution method in atmospheric escape simulation}

\footnote{\today}

\correspondingauthor{Yuichi Ito}
\email{yuichi.ito.kkyr@gmail.com}

\author[0000-0002-0598-3021]{Yuichi Ito}
\affiliation{Division of Science, 
National Astronomical Observatory of Japan
2-21-1 Osawa, Mitaka, Tokyo 181-8588, Japan }
\affiliation{Department of Physics and Astronomy, University College London 
Gower Street, WC1E 6BT London, United Kingdom}

\author[0009-0003-0399-1305]{Tatsuya Yoshida}
\affiliation{Faculty of Science, Tohoku University, 6-3 Aoba, Aramaki, Aoba-ku, Sendai 980-8578, Japan}

\author[0000-0002-0998-0434]{Akifumi Nakayama}
\affiliation{Department of Physics, College of Science, Rikkyo University, 3-34-1 Nishi-Ikebukuro, Toshima-ku, Tokyo 171-8501, Japan}

\begin{abstract}
 Atmospheric escape is crucial to understand the evolution of planets in and out of the Solar system and to interpret atmospheric observations. While hydrodynamic escape simulations have been actively developed incorporating detailed processes such as UV heating, chemical reactions, and radiative cooling, the radiative cooling by molecules has been treated as emission from selected lines or rotational/vibrational bands to reduce its numerical cost. However, ad hoc selections of radiative lines would risk estimating inaccurate cooling rates because important lines or wavelengths for atmospheric cooling depend on emitting conditions such as temperature and optical thickness. In this study, we apply the correlated-k distribution (CKD) method to cooling rate calculations for H$_2$-dominant transonic atmospheres containing H$_2$O or CO as radiative species, to investigate its numerical performance and the importance of considering all lines of the molecules. Our simulations demonstrate that the sum of weak lines, which provides only 1~\% of the line emission energy in total at optically thin conditions, can become the primary source of radiative cooling in optically thick regions, especially for H$_2$O-containing atmospheres.   Also, in our hydrodynamic simulations, the CKD method with a wavelength resolution of 1000 is found to be effective, allowing the calculation of escape rate and temperature profiles with acceptable numerical cost. Our results show the importance of treating all radiative lines and the usefulness of the CKD method in hydrodynamic escape simulations. It is particularly practical for heavy-element-enriched atmospheres considered in small exoplanets, including super-Earths, without any prior selections for effective lines.
\end{abstract}

\section{Introduction} \label{sec:intro}
Hydrodynamic escape of atmospheres is one of the important phenomena 
influencing the atmospheric/bulk composition and mass of planets highly irradiated by stellar X-ray and UV, especially for small planets \citep[e.g.,][]{Lammer+2008,Tian2015,Owen+2019}.
The discovery of numerous close-in exoplanets over the past two decades has spurred the active development of theoretical models for hydrodynamic escape in planetary atmospheres
\citep[see][and the references therein]{Owen+2019}.
Subsequent detailed hydrodynamic simulations have 
incorporated
radiative transfer and photo and thermal chemistry
\citep[e.g.,][]{Yelle2004,Garcia2007,Murray-Clay+2009,Koskinen+2013a,Koskinen+2013b}. 
However, 
 most of them focus on
 the hydrogen-dominant atmospheres, given the observed rapid escape of hydrogen and helium in hot Jupiters and sub-Neptunes
\citep[e.g.,][]{Vidal-Madjar+03,Kulow+2014,Spake+2018}.

Conversely, modeling the atmospheres enriched with heavy elements, which are likely in small-sized terrestrial planets including super-Earths, remains in its infancy.
Recent comparisons between observed absorption lines and theoretical modeling of the escaping atmosphere suggest the heavy-element-enriched atmosphere in the super-Earth of $\pi$ Men c \citep{Garcia+20,Garcia+21}.
Additionally, some planetary formation theories propose
that a primordial atmosphere would be naturally enriched with heavy elements \cite[e,g.,][]{Booth2019, Kimura2020}.
Thus,
modeling the hydrodynamic escape of such atmospheres becomes crucial, both for interpreting upcoming observations and understanding planetary evolution.

One of the key processes
in modeling
hydrodynamic escape is a radiative cooling.
This process removes thermal energy from atmospheres heated up by the high energy photons from its parent star, profoundly impacting both 
the thermal structure and the escape rate
\citep[e.g.,][]{Salz+2016}. 
In the case of atmospheres enriched in H$_2$O,  \citet{Yoshida+2022} newly considered the radiative cooling in the rotational-vibrational bands of H$_2$O
and found that it 
resulted in
the smaller escape rate with the higher mixing ratio of H$_2$O.
Such a strong cooling effect of H$_2$O was not found in another hydrodynamic simulation \citep{Johnstone2020} which considered only the rotational bands of H$_2$O as its cooling. 
The discrepancy can be explained by the more efficient cooling stemming from vibrational bands compared to rotational bands in hot escaping atmospheres. This is because vibrational transitions have higher excitation temperatures than those of rotational transitions \citep[][]{Yoshida+2022}.

Effective radiative lines for the cooling are
determined by 
emitting conditions such as the temperature and optical depth which are generally controlled by the altitude, atmospheric compositions, and more.
Therefore, when the specific structure of the atmosphere is unknown, it would be conservative to consider all lines.
However, there is a numerical difficulty in considering
all radiative lines of molecules. 
This is because the number of radiative lines of molecules is generally too large \citep[e.g., the number of H$_2$O lines is about half a million in {\it{HITRAN}} database,][]{GORDON+2022} to be handled in hydrodynamic simulations.

Consequently,
previous hydrodynamic simulations \citep[e.g.,][]{Johnstone2020, Garcia+21} have included 
only rotational lines in the ground vibrational state of H$_2$O using the analytical formula shown in \citet{Hollenbach+1979}.
Also, \citet{Yoshida+2022} considered 5000 lines of H$_2$O with their selection way of strong lines  \citep[][hereafter YK2020]{YK2020} but not all the lines. 
While such a line selection, as demonstrated by \citet{Johnstone2020} and \citet{Yoshida+2022}, can be an efficient approach to incorporating molecular lines in hydrodynamic simulations, it is hard to evaluate whether the neglected weak lines hardly contribute to a cooling rate or not.
This is because the weaker lines, which have smaller optical depth and a much larger number of lines than stronger ones, may significantly contribute to the cooling rate when stronger lines are optically thick.
Thus,
it is worth considering all lines of molecules but hydrodynamic simulations require an acceptable small numerical cost with maintaining accuracy.

To explore an effective way to consider all molecule lines in hydrodynamic simulations, 
this study examines
how efficiently a correlated-k distribution (CKD, hereafter) method \citep{Liou1980,Lacis+1991,Thomas+2002} works.
The CKD method is a grouping method of radiative lines and derives the absorption or emission probability density in wavelength bands which successfully reproduces the result of more sophisticated line-by-line calculations despite the small computational costs \citep[e.g.,][]{Irwin+2008,Amundsen+2017}.   
The method has been widely used in general circulation models and 1-D radiative transfer models for the lower atmosphere with high-pressure conditions, where the pressure broadening is effective, in solar-system planets and exoplanets \citep[e.g.,][]{Showman+2009,Wordsworth2013}.
While it hasn't been used in any simulations of hydrodynamic escape which are in low pressure conditions resulting the sparse and wide distributions of absorption lines 
because of ineffective broadening effects.

The remainder of this paper is organized as follows.
In Section.~\ref{sec:mod}, we describe the numerical setting to evaluate how the CKD method works in cooling rate calculations and hydrodynamic simulations.
In Section.~\ref{sec:res},
we show the comparison of cooling rates between the line selection (LS) method of YK2020 and the CKD method and the result of hydrodynamic simulation with the CKD method. 
In Section.~\ref{sec:disc}, we discuss the sensitivity of the resultant cooling rate to the numerical settings of the CKD method, the caveats of our investigation, the LS method and application to astronomical environments other than
atmospheres.
Finally, we summarize our results in Section~\ref{sec:conc}.

\section{Model} \label{sec:mod}
We assess 
how efficiently and accurately the CKD method works in the calculation for the hydrodynamic simulations of upper atmospheres.
For the assessment, we consider Mars-mass planets with H$_2$-dominant
atmospheres containing H$_2$O or CO as radiative species.
We chose these gases
because
the two gases can play one of the dominant coolants in upper atmospheres 
and hydrodynamic simulations with the LS method were previously shown \citep{YK2020,Yoshida+2022}.
Note that the planetary mass assumed in H$_2$-H$_2$O atmosphere in this study is different from that in \citet{Yoshida+2022}. 
Also, since the radiative lines of H$_2$O
densely and widely present
in infrared wavelength, while the radiative lines of CO 
are sparsely distributed 
in wavelength \citep[see][]{Tennyson+2018}, they would be good examples for demonstrating the versatility of the CKD method
from the viewpoint of their different radiative properties.
Although the effect of non-Local thermal equilibrium also plays a crucial role
in radiative properties \cite[e.g.,][]{Lopez2001}, YK2020 showed that emission from molecules dominates in the lower thermosphere where the pressure is sufficiently high to achieve local thermal equilibrium (LTE).
Thus, we also assume LTE conditions according to the YK2020 for our simulations and discuss the non-LTE effect in Sec.~\ref{sec:disc21}.

We consider two different transonic structures: one is the idealized isothermal atmospheres and 
the other is the non-isothermal atmospheres simulated by the YK2020 model.
For the isothermal transonic atmosphere, we assume homogeneous structures with the radiative species fractions (H$_2$O/H$_2$ or CO/H$_2$) of 0.01, 0.1 and 1, and temperatures  of 270~K, 500~K, 900~K, 1000~K, and 2000~K.
The vertical density profiles are estimated from the isothermal wind equations consisted of mass and momentum conservation
\citep[][]{Parker1964}. 
The lower and upper boundaries are set at 
$r=R_\mathrm{p}$ and $50R_\mathrm{p}$, respectively, where $r$ is the radial distance from the planetary center and $R_\mathrm{p}$ is the radius of the planet (=3390 km). The number density of H$_{2}$ at the lower boundary is assumed to be
$1\times 10^{13}\,\mathrm{cm^{-3}}$. 
The H$_{2}$O(CO)/H$_{2}$ ratio and temperature at the lower boundary are given as parameters. 

For the non-isothermal atmospheres, we 
adopt the YK2020 model. 
The model solves the fluid equations considering chemical and radiative processes for a multi-component gas assuming spherical symmetry to obtain the structure of steady outflows
\citep[please see YK2020 and][for the  details on the model]{Yoshida+2022}.
This model has been applied to simulate the hydrodynamic escape of H$_2$-CH$_4$-CO atmospheres for
early-Mars in YK2020 and of H$_2$-H$_2$O atmospheres for rocky exoplanets around M-type stars in \citet{Yoshida+2022}.
Especially for the radiative cooling calculation for this hydrodynamic simulation, we adopt not only the LS method of YK2020 but also the CKD method.
While YK2020 and \citet{Yoshida+2022} took into account not only H$_2$O and CO, but also other gases like CH$_4$ for radiative cooling, we simplify the cooling by excluding these other gases. 
This is done in order to discern the efficiency of the CKD method when considering the differing radiative properties between H$_2$O and CO.

As for the boundary conditions of non-isothermal atmospheres, 
 the lower and upper boundaries are set at $r=R_{p}+1000\mathrm{km}$ and 50$R_{p}$, respectively. The gas at the lower boundary is assumed to be composed of H$_{2}$ and H$_{2}$O (or CO) with 
 an H$_{2}$ number density of $1\times 10^{13}\,\mathrm{cm^{-3}}$ and the H$_{2}$O(CO)/H$_{2}$ ratio of 0.1. The temperature at the lower boundary is set at 150 K for the H$_{2}$-CO atmospheres \citep{YK2020} and 400 K for the H$_{2}$-H$_{2}$O atmospheres \citep[][]{Yoshida+2022}, respectively. Also, 
we adopt the stellar spectrum 
estimated for TRAPPIST-1 \citep[MODEL 2B in][]{Peacock+2019} for the H$_{2}$-H$_{2}$O atmospheres as with \cite{Yoshida+2022} and the solar spectrum estimated for the young Sun \citep[][]{Claire+2012} for the H$_{2}$-CO atmospheres as with \citep{YK2020}. 
\cite{Yoshida+2022} utilized the TRAPPIST-1 spectrum enhanced by a factor of 10 to 300, but we use the spectrum as is to discuss the effect of molecular radiative cooling which is more effective in the smaller XUV  condition.
Then, the assumed total energy fluxes for both atmospheres are similar with each other (1.78$\times10^2$ erg/cm$^2$s for the H$_{2}$-H$_{2}$O atmospheres and 2.42 $\times10^2$ erg/cm$^2$s for the H$_{2}$-CO atmospheres).
Note, the resultant chemical composition and the velocity profile are not presented in Section~\ref{sec:res}. This is because, for the isothermal atmospheres, they are simply derived from the isothermal wind equations \citep{Parker1964}. 
For the non-isothermal atmospheres, the profiles of number density and velocity have been already discussed
in \cite{YK2020} and \cite{Yoshida+2022} (please see these studies  for the details).
Also, we are discussing the effects of differences in planetary mass and XUV intensity from these studies  in the Appendix.

We calculate cooling rates from the atmospheres to space as same as the manner of 
hydrodynamic simulation of the YK2020. 
The radiative cooling rates, $Q_{\rm{rad}}$, are given by 
\begin{eqnarray}
Q_{\rm{rad}}= \int 2 \pi \sigma_\lambda B_\lambda(T) n_{s} \exp{(-\tau_\lambda)} d\lambda,
\label{eq:qrad}
\end{eqnarray}
where $T$ is temperature, $\sigma_\lambda$, $B_\lambda(T)$ and $\tau_\lambda$ are absorption cross section, Planck function and optical depth at wavelength, $\lambda$, respectively, and $n_{s}$ is the number density of gas species $s$. 
The optical depth is calculated by the vertical integration from a given altitude to the upper boundary in our numerical domain. 
We use not only the CKD and LS method of YK2020 model but also a line-by-line (LBL) method to compare with the cooling rates calculated from these methods. 
The LBL method is the most accurate method and thus used for estimating the error in the cooling rate calculations.  
Then, we consider the local maximum error and volume averaged error in the cooling rate, which are given by
\begin{eqnarray}
\epsilon_{max}&=& \max \left( \left| 1-Q_{\rm{rad}}'/Q_{\rm{LBL}} \right| \right),
\\
\epsilon_{ave}&=&  \int r^2 \left| 1-Q_{\rm{rad}}'/Q_{\rm{LBL}} \right| dr /  \int r^2dr,
\label{eq:err}
\end{eqnarray}
where $Q_{\rm{LBL}}$ and $Q_{\rm{rad}}'$ are $Q_{\rm{rad}}$ calculated with the LBL method and the others, respectively.

In the LS method, 
we use the same selected lines as 
YK 2020 for CO lines and \citet{Yoshida+2022} for H$_2$O lines.
The LS method contains 5021 transitions of H$_{2}$O and 638 transitions of CO referring to {{\it{HITRAN}}} database \citep[][]{Rothman+13} to cover more than 99 \% of the total energy emission under
the optically thin condition in the temperature range of 100--1000 K. Also, following YK2020, we calculate 
each line emission
by assuming the line shapes determined by Doppler-broadening for the radiative cooling calculations. 
Note that, in this study, we neglect the Doppler shift of the emission line in wavelength due to outflow acceleration considered by YK2020 since its effect on the radiative cooling rate is negligible in the subsonic region where the infrared thermal line emission occurs dominantly. We discuss the Doppler shift effect in Sec.~\ref{sec:disc22}.

In the CKD and LBL methods, we adopt open-access data of absorption cross sections provided by \citet{Molliere+2019}.
The line data on which the absorption cross sections of CO and H$_2$O are based is from the {\it{HITEMP}} database \citep{Rothman+2010} which has the larger number of line data  than the {\it{HITRAN}} database used in the LS method.
However, the difference of line database never affects our conclusions because the two databases make no difference in low temperature conditions of special interest in this study.
The data of absorption cross sections are provided 
in a numerical table in which high resolution ($\text{R}=\lambda/\Delta \lambda=10^6$) cross sections ranging from 0.3 to 28 micron in a wavelength region 
are given as functions of temperature and pressure
\citep{Molliere+2019}.
They assumed the Voigt profile for the line profile function, unlike the LS method.
However, the difference in the profile function never affects our results because we focused on low pressure conditions where the Voigt profile approaches the Doppler profile. 
As we consider upper atmospheres with lower pressure than 10$^{-5}$~bar, the absorption around line cores, which is mainly determined by the Doppler broadening and thus the temperature,  
is more important 
than the absorption at line wings 
determined by pressure broadening
in our results. 
Also, we have checked that the width of natural broadening is larger than the width of pressure broadening at 10$^{-5}$~bar
in most of the line data in the line database, {{\it{HITRAN}}} \citep{Rothman+13}, using the equation of natural broadening width given by \citet{Piskunov+01}. Therefore, we ignore the pressure dependence and adopt the data of absorption cross sections for 10$^{-5}$~bar.

In the CKD method, we compute a k-coefficient, which is opacity given by $\sigma_\lambda$ over mean atmospheric gas particle mass, table (hereafter, k-table) by dividing the overall spectrum range into wavelength bands for given temperature grids, sorting absorption coefficients in ascending order and then, transforming into a cumulative probability space with a given number of integration points, which is expressed as \textit{g}, for each band.
We use an open code of the CKD method, Exo\_k \citep{Leconte2021}, and the absorption cross-section data \citep{Molliere+2019} to make the k-table.
Exo\_k is an open-source Python library designed to handle many different formats of k-coefficient and cross-section tables in a computationally efficient way.
For nominal calculations, we set the temperature grids of the table to 
be the same as the grids of the cross-section data which are 81~K. 110~K, 148~K, 200~K, 270~K, 365~K, 493~K, 666~K, 900~K, 1215~K, 1641~K, 2217~K and 2295~K. Also, following  \citet{Leconte2021}, 20 Gauss-Legendre quadrature points are used in the cumulative probability space of absorption coefficients, \textit{g}-space, of all the k-table. Also, in Section~\ref{sec:disc1}, we discuss the dependence on the temperature grid and number of quadrature points.
We construct wavelength bands with a constant wavelength resolution. 
Then, in order to see which resolution is high enough to give an error smaller than the other uncertain sources in hydrodynamic escape simulations, we make the table with wavelength resolutions of 100, 300, 1000, 3000, and 10000. One of the other uncertain sources in hydrodynamic escape simulations is the stellar X-ray and UV heating rate with at least one order of magnitude uncertainty, since there is one-two orders of magnitude uncertainty in stellar X-ray and UV flux depending on each stellar evolution \citep[e.g.,][]{Johnstone+2021,Richey-Yowell+2019,Schneider+2018}. 
With the k-tables, we calculate $Q_{\rm{rad}}$ in the manner of the CKD method \citep[see][for the details]{Liou1980}.

\section{Result} \label{sec:res}

\begin{figure*}
 \begin{minipage}{0.5\textwidth}
    \begin{center}
        ($a$) CO/\rm{H$_2$}=0.1 \& $T=$270~K
  \includegraphics[width=\textwidth]{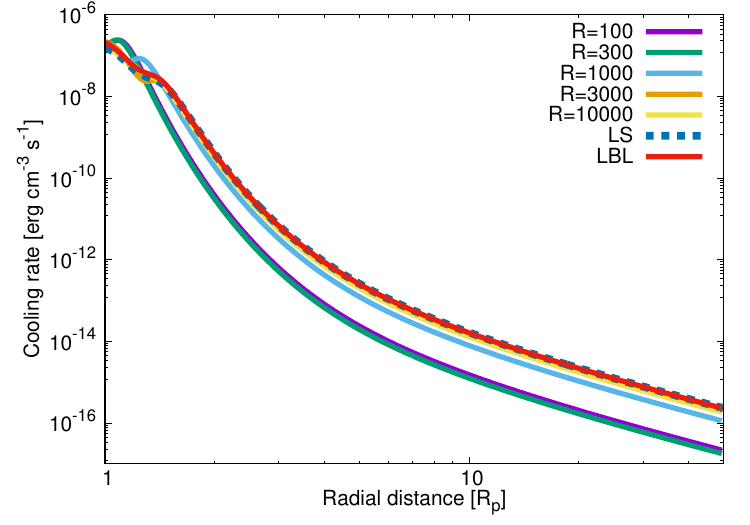}
   \end{center}
 \end{minipage}
   \begin{minipage}{0.5\textwidth}
    \begin{center}
        ($b$)  \rm{H$_2$O}/\rm{H$_2$}=0.1 \& $T=$270~K
  \includegraphics[width=\textwidth]{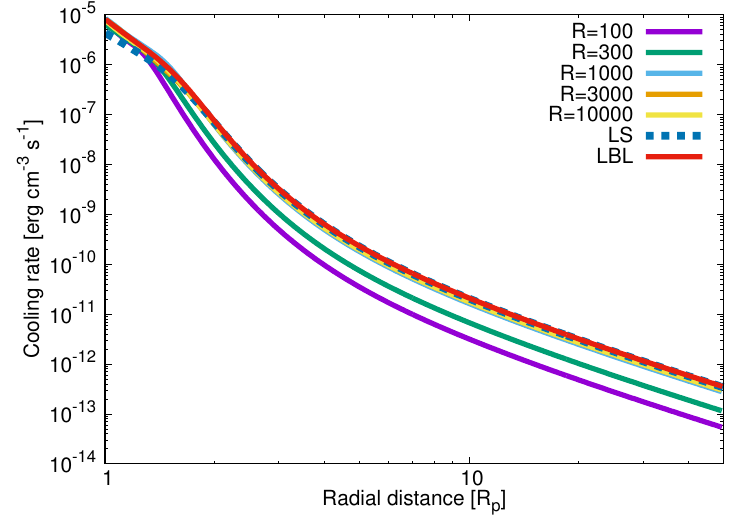}
   \end{center}
 \end{minipage}
  \begin{minipage}{0.5\textwidth}
    \begin{center}
    ($c$) CO/\rm{H$_2$}=0.1 \& $T=$1000~K
  \includegraphics[width=\textwidth]{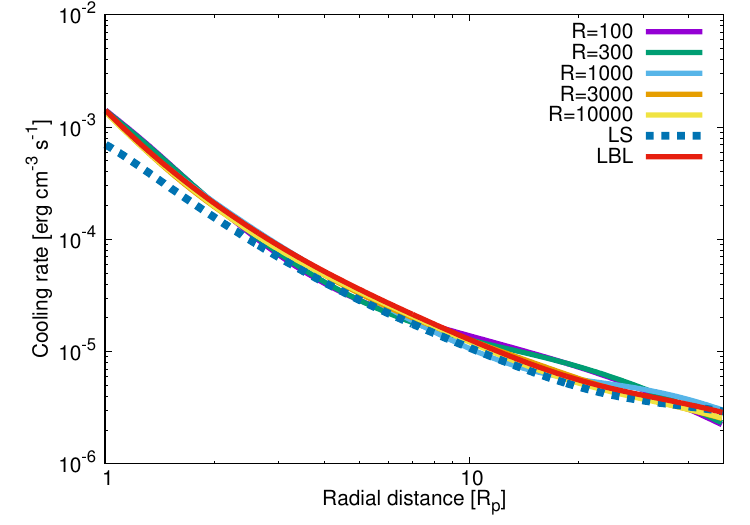}
   \end{center}
 \end{minipage}
   \begin{minipage}{0.5\textwidth}
    \begin{center}
            ($d$)  \rm{H$_2$O}/\rm{H$_2$}=0.1 \& $T=$1000~K
  \includegraphics[width=\textwidth]{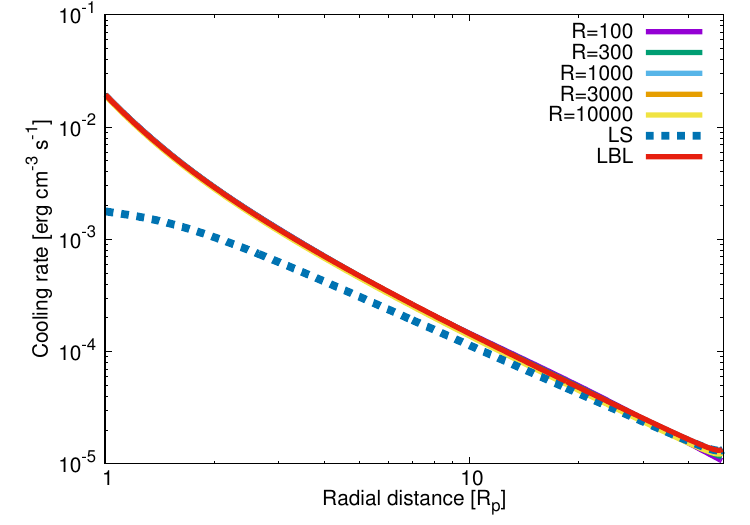}
   \end{center}
 \end{minipage} 

\caption{Comparison of cooling rate profiles with the CKD method for five choices of resolution $R$ of 100 (purple), 300 (green), 1000 (light blue), 3000 (orange) and 10000 (yellow), line selection (LS) method (dashed dark blue) and LBL method (red) for CO and \rm{H$_2$}O in \rm{H$_2$}-dominant
atmospheres. The horizontal axis is the radial distance from the planetary center in the unit of Martian radius. For the atmospheric properties, we assume different atmospheric composition and isothermal profiles:  ($a$) CO/\rm{H$_2$}=0.1 \& $T=$270~K, ($b$) \rm{H$_2$O}/\rm{H$_2$}=0.1 \& $T=$270~K, ($c$) CO/\rm{H$_2$}=0.1 \& $T=$1000~K, and ($d$) \rm{H$_2$O}/\rm{H$_2$}=0.1 \& $T=$1000~K.
}
\label{fig:atom_comp1}
\end{figure*}

\begin{figure*}
 \begin{minipage}{0.5\textwidth}
    \begin{center}  
     ($a$) $\rm{H_2}$+CO atmospheres
    \includegraphics[width=\textwidth]{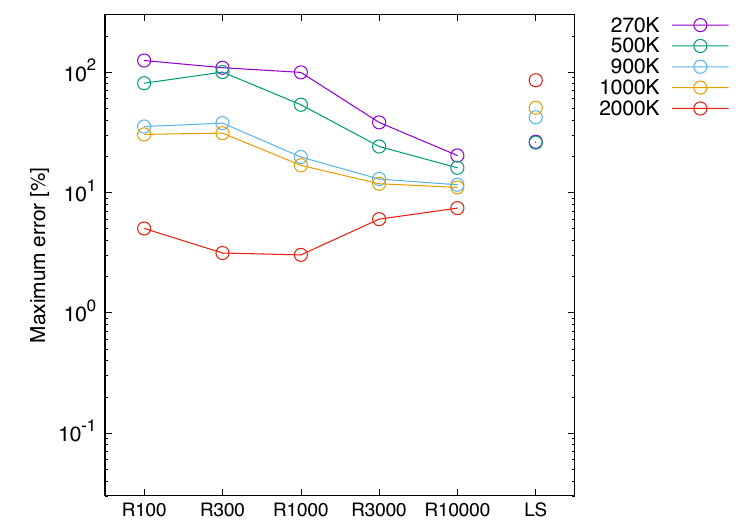}
   \end{center}
 \end{minipage}
   \begin{minipage}{0.5\textwidth}
    \begin{center}
    ($b$) $\rm{H_2}$+$\rm{H_2}$O atmospheres
    \includegraphics[width=\textwidth]{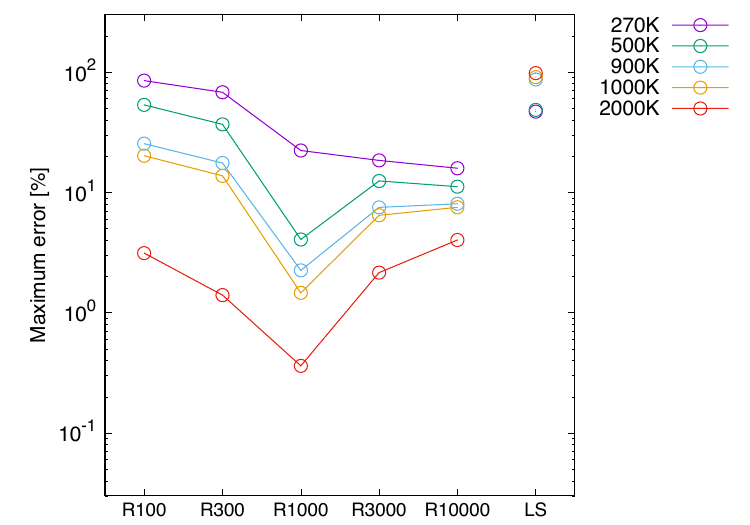}
   \end{center}
 \end{minipage}
 \begin{minipage}{0.5\textwidth}
    \begin{center}
  \includegraphics[width=\textwidth]{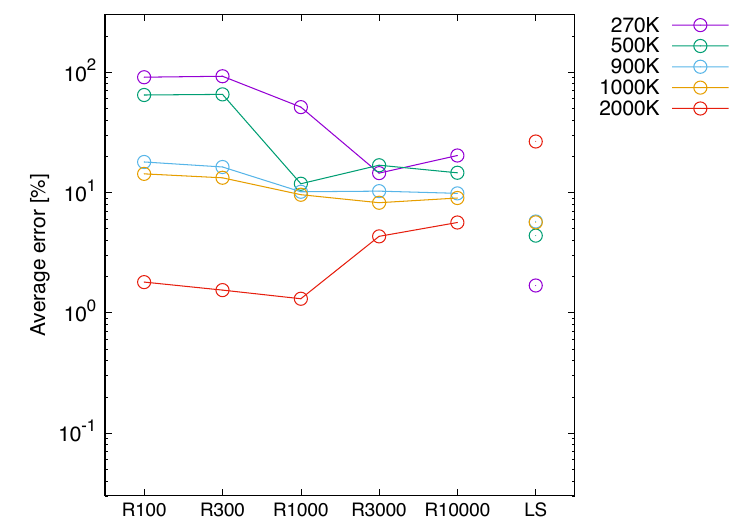}
   \end{center}
 \end{minipage}
  \begin{minipage}{0.5\textwidth}
    \begin{center}
  \includegraphics[width=\textwidth]{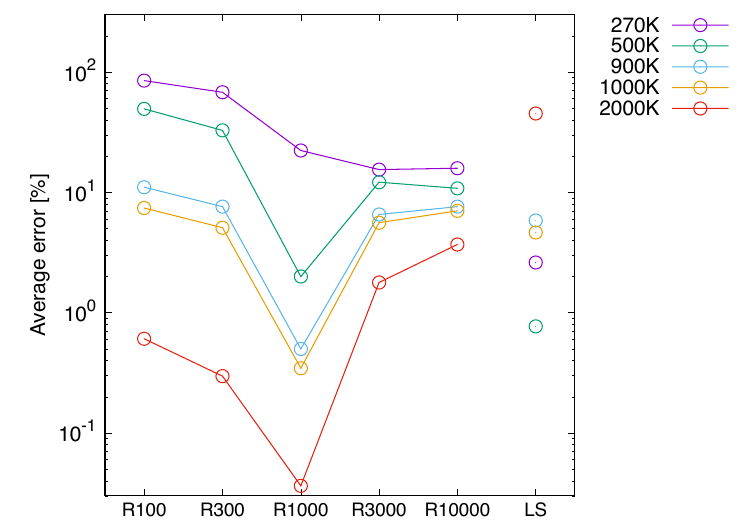}
   \end{center}
 \end{minipage}
\caption{Error in cooling rate calculations for CO ($a$) and \rm{H$_2$}O ($b$) in \rm{H$_2$}-dominant 
atmospheres with temperatures of 270~K (purple), 500~K (green), 900~K (light blue), 1000~K (orange) and 2000~K (red).  The CKD method for five choices of resolution of 100, 300, 1000, 3000, and 10000, and the LS method are used for the cooling rate calculations, which are shown in the horizontal axis.
Upper and lower panels respectively show the maximum error and the space-averaged error, while left-side and right-side panels respectively show the two errors for \rm{H$_2$}-dominant
atmospheres containing only CO and only $\rm{H_2}$O. 
In the atmospheres, the ratios of each two species (i.e., CO and \rm{H$_2$}O) and \rm{H$_2$}
are assumed to be 0.1.
}
\label{fig:atom_comp2}
\end{figure*}

\begin{figure*}
\begin{center}
 \begin{minipage}{0.45\textwidth}
      \begin{center}
 $\rm{H_2}$+CO atmospheres
 \end{center}
    \begin{center}  
    ($a$)
    \includegraphics[width=\textwidth]{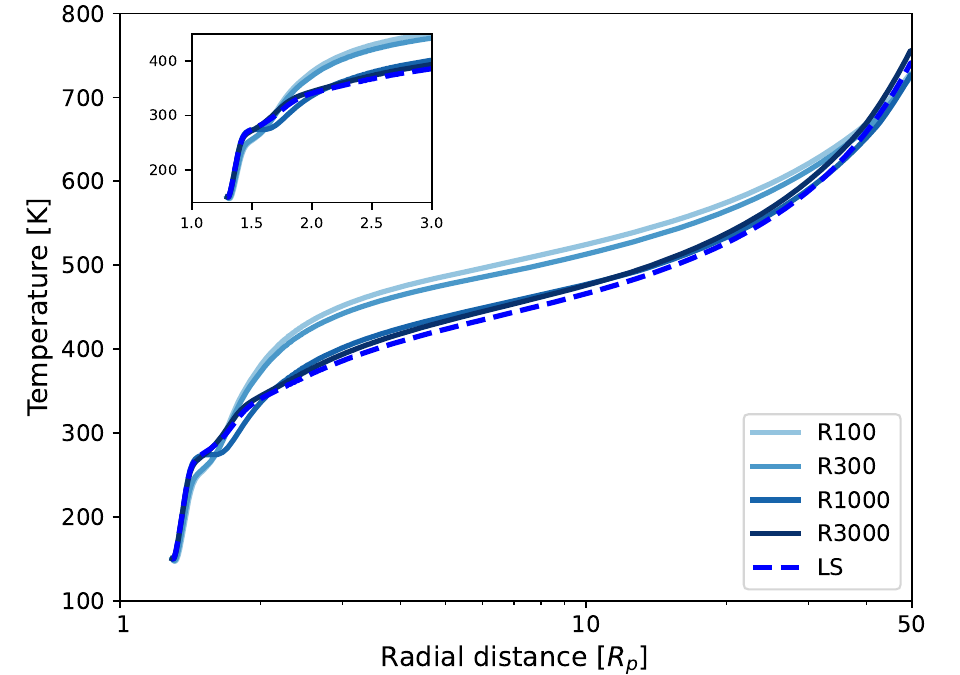}
   \end{center}
 \end{minipage}
   \begin{minipage}{0.45\textwidth}
          \begin{center}
$\rm{H_2}$+$\rm{H_2}$O atmospheres
 \end{center}
    \begin{center}
    ($b$)
    \includegraphics[width=\textwidth]{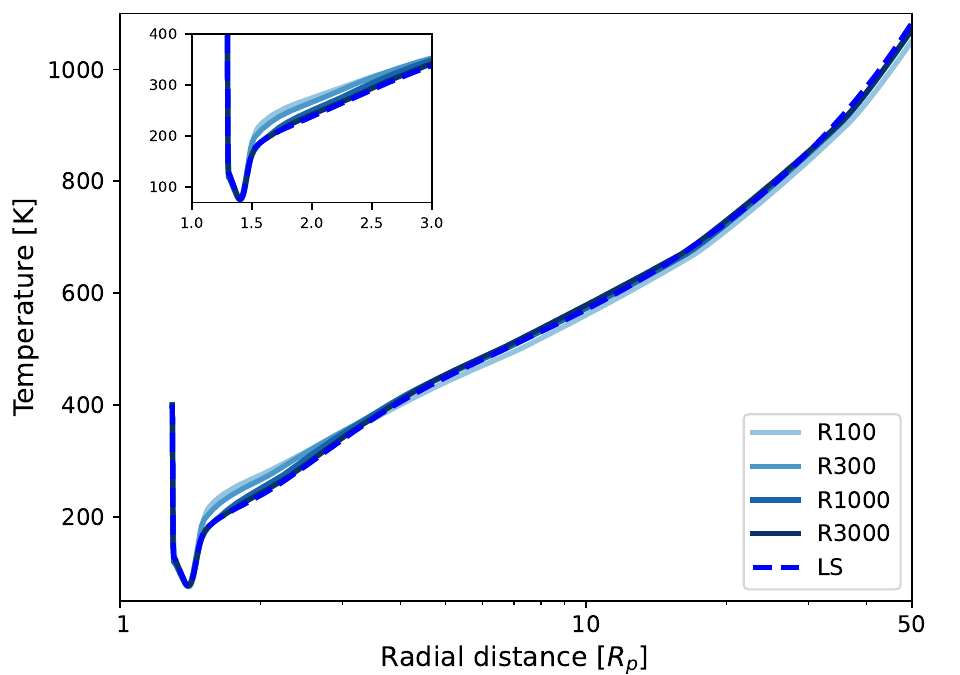}
   \end{center}
 \end{minipage}
 \\
 \begin{minipage}{0.45\textwidth}
    \begin{center}
    ($c$)
  \includegraphics[width=\textwidth]{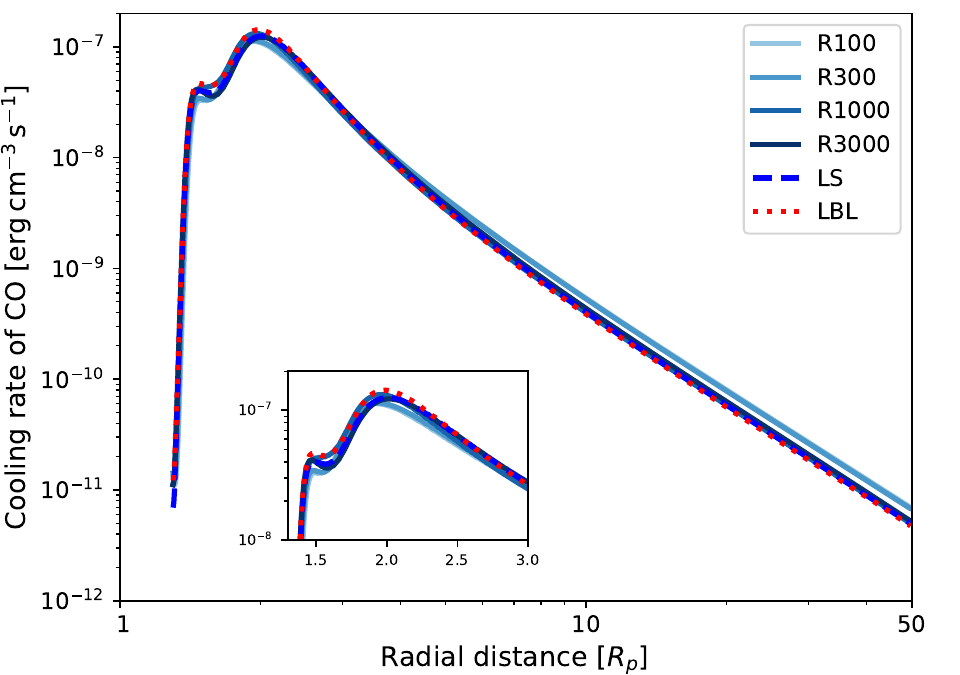}
   \end{center}
 \end{minipage}
  \begin{minipage}{0.45\textwidth}
    \begin{center}
    ($d$)
  \includegraphics[width=\textwidth]{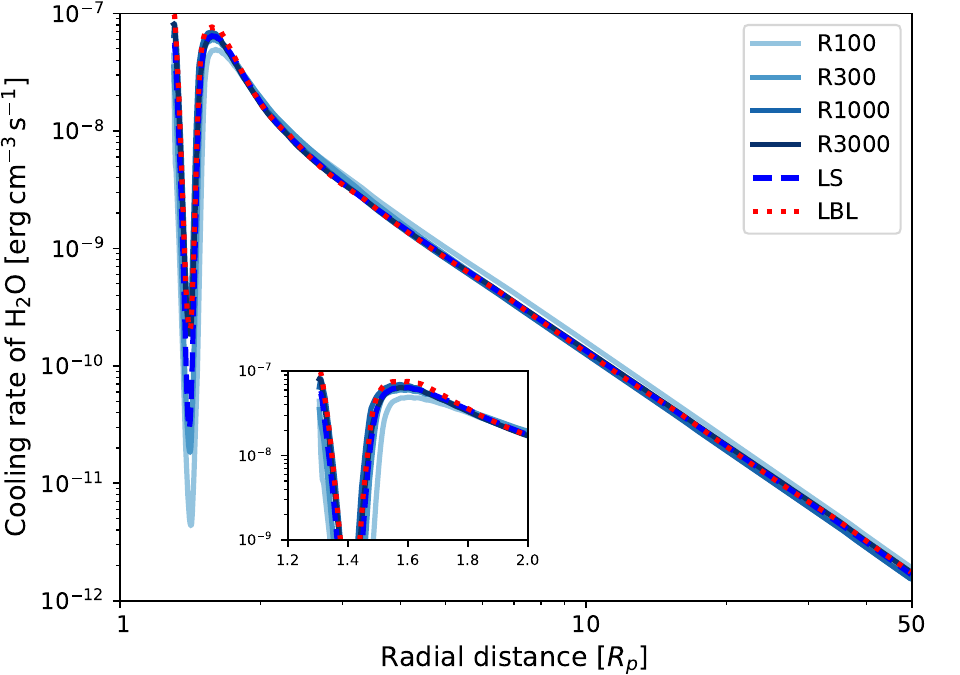}
   \end{center}
 \end{minipage}
  \\
  \begin{minipage}{0.45\textwidth}
    \begin{center}
    ($e$)
  \includegraphics[width=\textwidth]{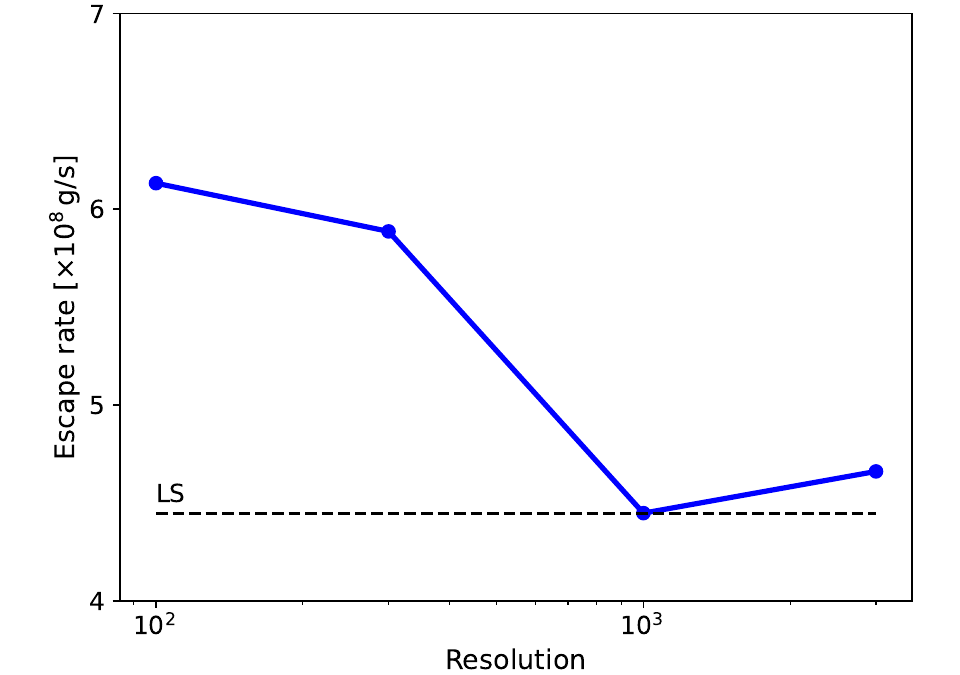}
   \end{center}
 \end{minipage}
  \begin{minipage}{0.45\textwidth}
    \begin{center}
    ($f$)
  \includegraphics[width=\textwidth]{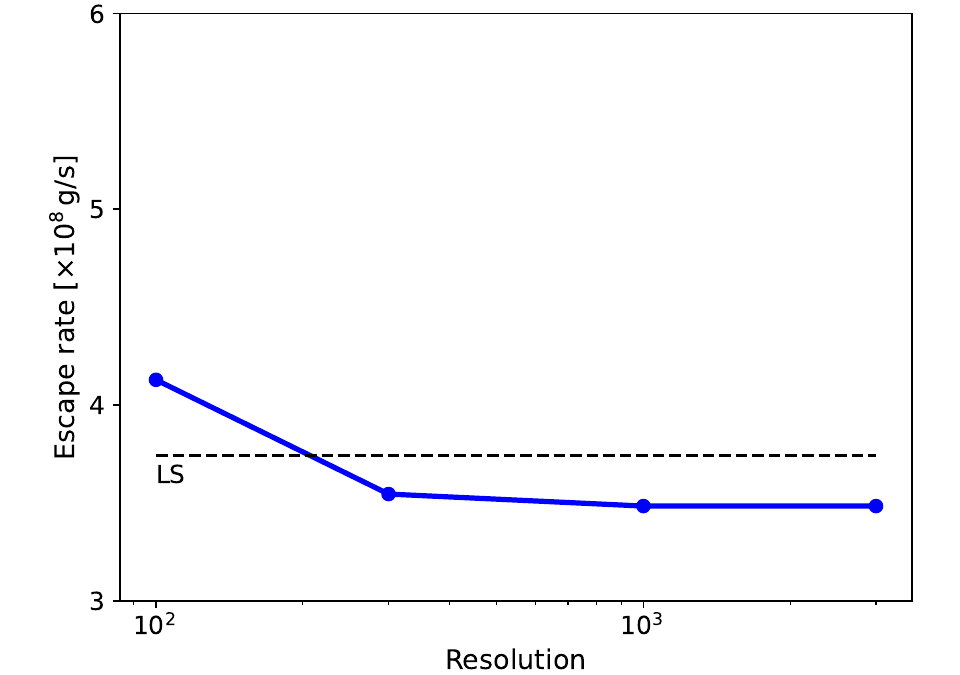}
   \end{center}
 \end{minipage}
 \end{center}
\caption{Calculated temperature profiles, cooling rate profiles, and escape rates of the non-isothermal H$_{2}$-CO atmospheres (left column) and H$_{2}$-H$_{2}$O atmospheres (right column) obtained by applying the CKD method with a wavelength resolution, R, of 100, 300, 1000, and 3000 to calculate the cooling rate of CO and H$_{2}$O. 
(a) and (b): Temperature profiles of the H$_{2}$-CO atmospheres and H$_{2}$-H$_{2}$O atmospheres, respectively. 
(c) and (d): Cooling rate profiles in the H$_{2}$-CO atmospheres and H$_{2}$-H$_{2}$O atmospheres, respectively. 
For reference, we also show the profiles obtained by the LS method and the LBL method. Note that the radiative cooling rate of the LBL method is estimated  by  the atmospheric profile as same as that of the LS method.
(e) and (f): Escape rate as a function of the resolution R for the H$_{2}$-CO atmospheres and the H$_{2}$-H$_{2}$O atmospheres, respectively.
The dashed line represents the escape rate obtained by applying the LS method. The profiles of gas species and velocity for  the CKD method with R of 1000 are shown in Fig~\ref{fig:num} at Appendix. 
Note that, a fluid approximation in the model is invalid above 40~R$_p$ which is exobase for these two atmospheres.
}
\label{fig:calc_escape}
\end{figure*}

Here, we investigate the radiative cooling rates of CO and H$_2$O. 
As described in the Introduction, the focus of this study is the accuracy of the CKD method in the condition of hydrodynamic upper atmospheres. 
Furthermore, we derive the difference
in estimates of escape rates and temperature structure involved in radiative cooling methods.

Firstly, we demonstrate the accuracy of the CKD method for isothermal atmospheres with different temperatures and compositions.
Fig.~\ref{fig:atom_comp1} shows the calculated cooling rate profiles of CO and \rm{H$_2$O} from the CKD method with R of 100 (purple), 300 (green), 1000 (light blue), 3000 (orange), and 10000 (yellow), the LS method (dashed dark blue), and the LBL method (red). 
Each panel of Fig.~\ref{fig:atom_comp1} shows these for 
($a$) CO/\rm{H$_2$}=0.1 \& $T=$270~K, ($b$) \rm{H$_2$O}/\rm{H$_2$}=0.1 \& $T=$270~K, ($c$) CO/\rm{H$_2$}=0.1 \& $T=$1000~K, and ($d$) \rm{H$_2$O}/\rm{H$_2$}=0.1 \& $T=$1000~K.

First, in the case of CO with 270~K (Fig.~\ref{fig:atom_comp1}$a$), the CKD method agrees with the LBL method within an order of magnitude for R$=1000$ and shows very good agreement with the LBL method for R$\geq3000$,
while there is approximately an order of magnitude difference in cooling rates between the CKD method with R~$\leq 300$ and the LBL method.
For the smaller R, substantial errors are found in both
smaller and larger radial distances.
Also, in the case of \rm{H$_2$O} with 270~K (Fig.~\ref{fig:atom_comp1}$b$), general trend is similar to the case of CO-enriched atmosphere.
However, the error of the CKD method in H$_2$O-enriched  atmospheres is smaller than that in CO-enriched  atmospheres and there is a small error in the smaller radial distance even if the smaller R.

Such error features come from grouping radiative lines in the CKD method.
In the CKD method, absorption cross-sections are integrated over the frequency of absorption strength in each wavelength bin.
Thus, the CKD method with the smaller R crudely evaluates the effects of a small number of absorption strengths, which are primarily strong and/or weak radiative lines.
Especially, when the number of strong and/or weak lines is small in each wavelength bin, the CKD method tends to be less accurate in an optically thinner and/or thicker region (i.e., larger and/or smaller radial distance),  respectively.
As a result, substantial errors are found in the smaller and larger radial distance in CO-enriched  atmospheres. 
By contrast, the CKD method in H$_2$O-enriched  atmospheres accurately estimates the cooling rates in the smaller radial distance because H$_2$O has a lot of weak lines. 
On the other hand,
the LS method accurately estimates the cooling rates everywhere compared with the CKD method with the smaller R.
This is because the LS method is close to the LBL method in optically thin conditions such as low temperature conditions which result in the small column density of radiatively active species. 

In the higher temperature (Fig.~\ref{fig:atom_comp1}$c$ and $d$), the differences between the CKD method and the LBL method become much smaller than those obtained in low temperature cases.
Furthermore, especially for the \rm{H$_2$O}-enriched  atmospheres, the CKD method agrees with the LBL method regardless of the resolution.
This is because efficient Doppler broadening and the contributions of weak lines in highly excited states due to high temperature \citep[e.g.,][]{Tennyson+2018} reduce the difference in absorption strengths within each wavelength bin, leading to the high accuracy of the integration in the CKD method.
On the other hand, in the H$_2$O-enriched atmospheres, the LS method is less consistent with the LBL calculation at the smaller radial distances since weak emission lines, neglected in the LS method, become relatively stronger. 
Then, the sum of the weak lines contributes to cooling rates in such small radial distances where strong emission lines become optically thick and thus inefficiently contribute to cooling rates.
On the contrary, in the CO-enriched  atmospheres, weak lines have a small contribution to cooling rates because CO has a small number of weak lines even in highly excited states \cite[e.g.,][]{Rothman+13}.

Figure \ref{fig:atom_comp2} shows the maximum and space-averaged error, $\epsilon_{max}$ and $\epsilon_{ave}$, in the cooling rate calculations for CO and \rm{H$_2$}O in \rm{H$_2$}-dominated atmospheres with the radiative species fraction of 0.1 and various temperatures.
$\epsilon_{max}$ of the CKD method for isothermal atmospheres are less than 100~\%, except for the cooling rates of CO with R of $\leq$ 300 and T $\leq$ 500~K.
$\epsilon_{max}$ of the CKD method usually decreases with R. 
However, in some cases, $\epsilon_{max}$ increases with R for $R \geq 1000$. 
For instance, $\epsilon_{max}$ monotonically decreases with R up to 10000  for CO with $T=$270-1000~K and  for \rm{H$_2$O} with $T=$270~K.
On the other hand, the error has the smallest values at R$=$1000 for CO with $T=2000$~K and for H$_2$O with $T\geq500$~K. 
This is because the CKD method integrates the line intensity in \textit{g}-space assuming that the cumulative distribution function of the intensity is a smooth and monotonically increasing function.
Since larger R reduces the number of lines in each wavelength bin, larger R inaccurately estimates the averaged intensity due to the dispersion of line intensity over \textit{g}-space although larger R provides better wavelength resolution of emission.
We confirmed such trends by the calculation of the error in an optically thin condition 
(
see Figure~\ref{fig:atom_comp3} in the Appendix). 
Also, as shown in Fig.~\ref{fig:atom_comp2},
$\epsilon_{ave}$ of the CKD method has almost the same trend over R with that of $\epsilon_{max}$.
This is because the space-averaged error size is susceptible to errors at large radial distances (see Eq.~\ref{eq:err}), and moreover the CKD method usually provides $\epsilon_{max}$ at large radial distances in our simulations.

  In the overall trend for all wavelength resolution, the CKD method works better for \rm{H$_2$}O than that for CO.
This is because \rm{H$_2$}O has more lines with similar strength than CO and thus the averaged values at each bin for \rm{H$_2$}O are more accurate than those for CO.
For the same reason, the CKD method works better for higher temperatures as the number of emission lines with similar strength increase with temperature, as shown in Fig.~\ref{fig:atom_comp2}.
Inversely, the LS method usually works worse for higher temperatures since it neglects weak emission lines that can affect the cooling rates in lower altitudes, shown in Fig.~\ref{fig:atom_comp1}.
Also, in the cases for the radiative species fraction of 0.01 and 1, $\epsilon_{max}$ is almost the same as that for the fraction of 0.1 and $\epsilon_{ave}$ is scattered a little bit, as 
shown in Fig.~\ref{fig:atom_comp4} in the Appendix. 
Then, the same overall dependence on R and the similar difference between CO and \rm{H$_2$}O are seen.

Figures \ref{fig:calc_escape} show the temperature profile, cooling rate profile, and escape rate of the non-isothermal H$_{2}$-CO and H$_{2}$-H$_{2}$O atmospheres with 10~\% of CO(H$_{2}$O) obtained by applying the CKD method and LS method. 
In these conditions, the total mass escape rate is about 
a factor of 5 and 10 lower than that of the pure hydrogen atmosphere for the H$_{2}$-H$_{2}$O atmosphere and H$_{2}$-CO atmosphere respectively, due to the radiative cooling effect by CO and H$_{2}$O \citep{YK2020,Yoshida+2022}. 
In the case of the H$_{2}$-CO atmosphere, the overall temperature with R$\leq 300$ is relatively high compared to that with the higher R (Fig.~\ref{fig:calc_escape}$a$) due to the error in the cooling rate calculation in this temperature range, as shown in the isothermal low-temperature cases with $T=270$ and $500$~K (see Fig.~\ref{fig:atom_comp1} and \ref{fig:atom_comp2}).
On the other hand, the difference in the profile of the temperature over R for the H$_{2}$-H$_{2}$O atmosphere is relatively small (Fig.~\ref{fig:calc_escape}$b$). 
The reasons for the difference in the dependence on R between the H$_{2}$-CO atmosphere and the 
 H$_{2}$-H$_{2}$O atmosphere are as follows.
Firstly, the CKD method can accurately estimate the H$_{2}$O cooling rate even when R is small due to the high affinity of this method with the characteristics of H$_{2}$O line emission as described in the previous paragraphs and Figures~\ref{fig:atom_comp1} and \ref{fig:atom_comp2}. 
Moreover, the H$_{2}$O cooling rate is sensitive to a change in temperature (Fig.~\ref{fig:dcool_comp} in the Appendix), which also contributes to the adjustment of the cooling rate by a slight change in temperature.
As a result of the less dependence of the temperature profile on R in the H$_{2}$-H$_{2}$O atmosphere, the difference in the cooling rate and escape rate of the H$_{2}$-H$_{2}$O atmosphere with R is also small compared with that of the H$_{2}$-CO atmosphere (Fig.~\ref{fig:calc_escape}$c,d,e,f$).

Compared with the result obtained by the LS method, the temperature is high, and the escape rate is large in the H$_{2}$-CO atmosphere when R$\leq 300$ (Fig.~\ref{fig:calc_escape}$a,e$) due to the underestimation of the CO cooling rate. 
On the other hand, the escape rate of the H$_{2}$-H$_{2}$O atmosphere when R$\geq 300$ is smaller than that with the LS method (Fig.~\ref{fig:calc_escape}$f$) since weak H$_{2}$O emission lines neglected by the LS method works as effective radiative cooling sources in the lower region of $r\lesssim 2R_{p}$ where strong emission lines become optically thick.
Such dependences are also realized by the cooling rate profiles for isothermal cases and the error of the LS method in the cooling rate calculation (see Fig.~\ref{fig:atom_comp1} and ~\ref{fig:atom_comp2}).

From our results, we found the CKD
 method with R$=1000$ is effective for estimating the escape rate and temperature profile.
In the cases with R$=1000$ for both the H$_{2}$-CO and H$_{2}$-H$_{2}$O atmospheres, the profiles of temperature and radiative cooling rate are in good agreement with the higher resolution case with R$=3000$ (Fig.~\ref{fig:calc_escape}$a,b,c,d$). Also, for both the atmospheres, the difference between the escape rates with R$=1000$ and that with R$=3000$ is only about 10~\% (Fig.~\ref{fig:calc_escape}$e,f$) and the radiative cooling rates are consistent with those obtained by the LBL method (Fig.~\ref{fig:calc_escape}$c,d$).
Note that the radiative cooling rate by the LBL method is obtained based on the atmospheric profile obtained by applying the LS method.
Whether the errors related to R in atmospheric profile and escape rate are allowed or not depends on the objective. 
Therefore, conservatively, the CKD with R=1000 is accurate enough to simulate not only mass loss rate but also temperature profile.

The CKD method can save computation time to obtain the radiative cooling rate by comprehensively considering numerous emission lines. 
As shown in Table \ref{tab:cost}, the computation time of the CKD method is shorter than that of the LS method when R $\leq 300$ 
for the H$_{2}$-H$_{2}$O atmosphere. 
On the other hand, it is longer than that of the LS method in all the cases of the considered R for the H$_{2}$-CO atmosphere. 
In terms of the computation time, the CKD method is especially effective when dealing with radiatively active species with a lot of emission lines such as H$_{2}$O.

\begin{table}
	\centering
	\caption{Computation time of radiative cooling  calculation with the CKR method and the LS method relative to that with the LBL method.}
	\label{tab:example_table}
 	\begin{tabular}{ccccccc} 
	R100 & R300 & R1000 & R3000 & LS(CO) & LS(H$_{2}$O) \\
    \hline
     0.0013 & 0.0060 & 0.032 & 0.13 & $5.1\times10^{-4}$ & 0.010 \\
	\end{tabular}
 \tablecomments{These are shown as the relative values of computation time when that with the LBL method is set to 1.}
\label{tab:cost}
\end{table}

\section{Discussion} \label{sec:disc}
\subsection{Dependence of grid points in k-table}
\label{sec:disc1}

In the results, we focus on the wavelength resolutions, R, in the CKD method. 
The numbers of grid points in k-table along with
not only R but also temperature-space and 
integration points in \textit{g}-space
are important to provide accurate calculations.
We adopt 20 points in \textit{g}-space and 13 points in temperature-space, which are respectively as same as the default integration points in \textit{g}-space (\textit{g}-grids, hereafter)
in Exo\_k \citep{Leconte2021} and the T-points of the cross section data \citep{Molliere+2019}, for the k-table.
Below, we discuss the sensitivity of the results to the grid points in \textit{g}-space and temperature-space.

\begin{figure}
    \begin{center}
\includegraphics[width=\columnwidth]{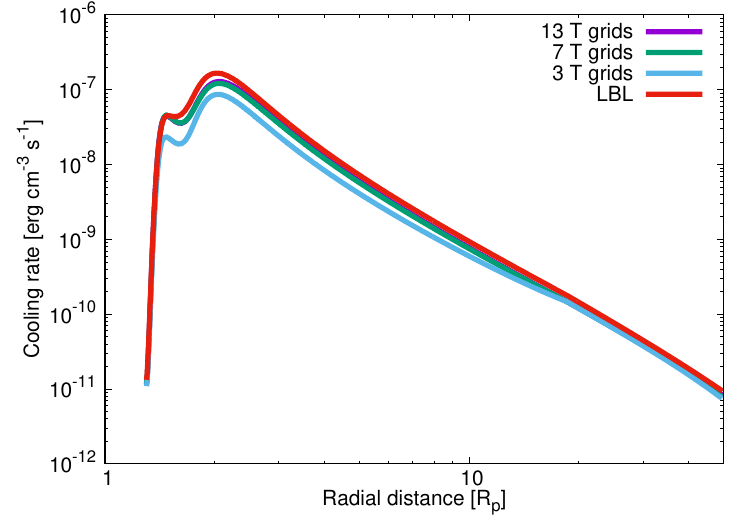}
\caption{Comparison of cooling rate profiles with the CKD method using a resolution of 3000 for three choices of temperature grid of 13 (purple), 7 (green), and 3 (light blue) and the LBL method  (red) for CO in \rm{H$_2$}-dominant
atmospheres. For the atmospheric properties, we used the atmosphere simulated by the LS method shown in Fig.\ref{fig:calc_escape}.
}
\label{fig:atom_comp_d1}
   \end{center}
\end{figure}

Figure~\ref{fig:atom_comp_d1} shows calculated cooling rate profiles of
CO from the CKD method with R of 3000 and different temperature grids for the atmosphere simulated by the LS method shown in Fig.\ref{fig:calc_escape}~$a$. 
For this sensitivity test, we used 13 temperature grids (purple), which is  the same as the one
we used in Sec.\ref{sec:res}, 7 temperature grids (green), which are 81~K, 148~K,  270~K,  493~K,  900~K,  1641~K, and 2295~K, and 3 temperature grids (light blue), which are 81~K, 493~K,  and 2295~K. 
In the case of the smallest temperature grids, as expected, the difference from the LBL calculation becomes larger than the others. 
On the other hand, the cooling rate profile for 7 temperature grids is almost same as that for the largest temperature grids. It suggests that the resolution of the temperature grid is  high enough to ensure that it is not the primary source of error.
Actually, the maximum error for 7 and 13 temperature grids are 35~\% and 33~\%, respectively, but the smallest temperature grid gives about two times larger error, 61~\%.
Also, we did this sensitivity test for R of 1000 and then found 
an almost same trend with the result for R of 3000.
Although the minimum resolution of the temperature grid, which is not to be the primary source of error, would depend on temperature structure and coolant species, this sensitivity test suggests that the resolution more than a doubling interval is suitable in the temperature range.

\begin{figure}
    \begin{center}
\includegraphics[width=\columnwidth]{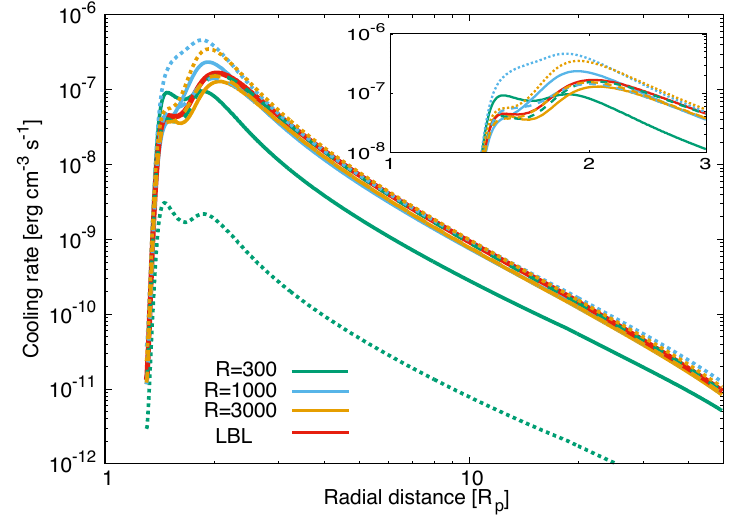}
\caption{Comparison of cooling rate profiles with the CKD method for three choices of {\textit{g}}-grid of 10 (dotted), 20 (solid), and 40 (dashed), and also three choices of resolution of 300 (green), 1000 (light blue), 3000 (orage), and the LBL method (red) for CO in \rm{H$_2$}-dominant
atmospheres.  For the atmospheric properties, we used the atmosphere simulated by the LS method shown in Fig.\ref{fig:calc_escape}. The inset is an enlarged view of the cooling rate profiles for
a radial distance from 1 $R_p$ to 3 $R_p$.
}
\label{fig:atom_comp_d2}
   \end{center}
\end{figure}

Figure~\ref{fig:atom_comp_d2} shows calculated cooling rate profiles of
CO from the CKD method with R of
300 (green), 1000 (light blue), 3000 (orange) and the LBL calculation (red), assuming the atmosphere simulated by the LS method shown in Fig.\ref{fig:calc_escape}~$a$.
Solid lines, dashed lines, and dotted lines show those with 10, 20, and 40
\textit{g}-grids, respectively.
The CKD method with smaller {\textit{g}}-grids gives less accurate values in the cooling rate profiles. 
Also, the dependence on the number of \textit{g}-grid 
becomes larger for smaller R. 
For instance, the difference 
between each number of {\textit{g}}-grid is about two orders of magnitude for R$=300$ while it becomes less than one order of magnitude for R$=1000$.
This is because the value of R and also the {\textit{g}}-grid number are key parameters for the k-table to retain the wavelength dependence information of the original absorption cross section. 

\subsection{Caveats of investigation} 
\label{sec:disc2}
\subsubsection{Non-LTE effect}
\label{sec:disc21}
Molecules in upper atmospheres are thought to often be in non-LTE conditions and then their cooling rates becomes less efficient than those in LTE ones \citep[e.g.,][]{Lopez2001}. 
In principle,
the non-LTE effect in the CKD method can be considered with k-table made from cross section data including the non-LTE effect by different collisional-/radiational-transition rates.
However, to our knowledge, the accessible collisional-transition data of multiatomic molecules in open database and published articles is limited. 
For instance, Leiden Atomic and Molecular Database,  
{\textit{LAMDA}}\footnote{https://home.strw.leidenuniv.nl/~moldata/},
currently provides the collisional-transition data of 47 polyatomic molecules but all the data is available only for collision with H$_2$, He or electrons \citep[][]{vanderTak+2020} and 
the provided collisional transition data is much smaller dataset than the radiative transition data provided in {{\it{HITRAN}}} database. 
Due to the data limitation, at present, it is difficult to take the non-LTE effect of many lines into account. 

Both the non-LTE effect and optical thickness make cooling rates inefficient. 
In general, the former effect is dominated at a lower pressure region and the latter one is dominated at a higher pressure region. 
To roughly estimate the non-LTE effect, we evaluate the escape rate of the H$_{2}$-H$_{2}$O atmosphere on the assumption of both non-LTE and LTE conditions by applying the formula derived by \citet{Hollenbach+1979}, which considers cooling in the water vapor rotation lines on the ground vibrational state 
in the non-LTE condition. 
The limitation of radiative cooling by non-LTE effects leads to an increase in the escape rate by a factor of $\sim 2$. 
It is one of the issues to develop a scheme to incorporate the non-LTE effects in order to estimate the escape rate and atmospheric profile more accurately.

\subsubsection{Velocity shift}
\label{sec:disc22}
\begin{figure}
    \begin{center}
\includegraphics[width=\columnwidth]{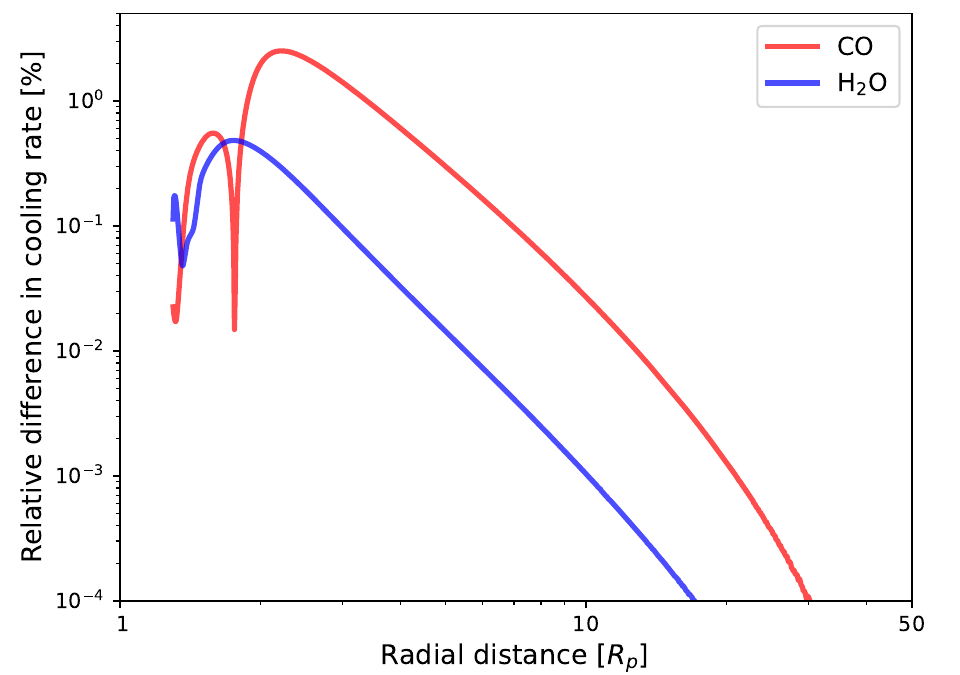}
\caption{Relative difference of the cooling rate of CO (red) and H$_{2}$O (blue) with the Doppler shift effect from that without the Doppler shift effect in the non-isothermal H$_{2}$-CO and H$_{2}$-H$_{2}$O atmospheres with CO(H$_{2}$O)/H$_{2}$=0.1.
}
\label{fig:cool_comp_velshift}
   \end{center}
\end{figure}

As gases move to space with a high velocity in hydrodynamic escaping atmospheres, their radiative lines should be shifted due to the Doppler effect \citep[e.g.,][]{Rybicki+1986}. 
If the velocity of gases becomes high enough to shift lines at optically thick regions, 
the Doppler shift effect  
should reduce the optical thickness and then increase the radiative cooling rate. 
However, as an atmospheric region with a velocity high enough to shift lines is at high altitudes around a sonic point, the lines of molecules would  usually be optically thin there. 
In this case, the Doppler shift hardly affects the optical thickness and cooling rate.
As an example, Figure \ref{fig:cool_comp_velshift} shows the relative difference of the cooling rate of CO and H$_{2}$O with the Doppler shift effect from that without the Doppler shift effect on the profiles of the non-isothermal H$_{2}$-CO and H$_{2}$-H$_{2}$O atmospheres with CO(H$_{2}$O)/H$_{2}$=0.1 obtained by applying the LS method. 
To estimate cooling rates, we apply the LS method with the model of \citet{YK2020} and \citet{Yoshida+2022}.
The difference is quite small within several \% in the calculation region, which shows the effect of the Doppler shift is negligible on the conditions considered by this study where the gaseous radial velocity is small compared with the sonic velocity in the lower optically thick region.
While we haven't 
explicitly include the effect in this study, it would be possible to take the effect into account by making k-tables incorporating the Doppler shift with different velocities. 

\subsubsection{Mixture of multi radiatively active species}
\label{sec:disc23}
The numerical techniques of the CKD method for modeling of overlapping absorption by multi-gas species has been proposed \citep[e.g.,][]{Lacis+1991,Goody+1989,Edwards+1996}. 
 According to \citet{Amundsen+2017} which investigated the accuracy and numerical cost of the CKD method for hot Jupiters atmospheres with several gases, the random overlap method \citep{Lacis+1991} was the most accurate and flexible method for one-dimensional models in their investigation and the equivalent extinction method \citep{Edwards+1996} was less accurate but could be an efficient method for three-dimensional models with a smaller numerical cost than the random overlap method. 
The random overlap and equivalent extinction methods are also applicable to 
the hydrodynamic atmosphere in principle and can be easily implemented into the CKD method.

Current hydrodynamic simulations suggest the importance of radiative cooling from many species: monoatomic gases such as H, O, N, Si, Mg and Na \citep[e.g.,][]{Murray-Clay+2009,Nakayama+2022,Ito+2021}; reducing gases such as H$_{3}^{+}$, CO and  CH$_4$ \citep[e.g.,][]{Yelle2004, YK2020, Yoshida+2021}; and oxidizing gases such as CO$_2$ and H$_2$O \citep[e.g.,][]{Johnstone+2018,Johnstone2020,Yoshida+2022}. Some of these species probably can coexist in the upper atmospheres of planets depending on X-ray and UV irradiation environments and atmospheric element abundance, and then their radiative lines may  overlap. In such a case, the CKD method could be an effective scheme to apply to a realistic atmosphere, modeling  overlapping absorption by such radiatively active species.
A detailed investigation of the performance of the CKD method for a mixture of radiatively active species is beyond the scope of this study and will be done in our future study.

\subsection{Line selection method}
\label{sec:disc3}
The LS method is useful in cooling rate calculations but the selection way needs to handle a trade-off between its accuracy and its numerical cost.
In Sec.~\ref{sec:res},
we showed how the LS method worked by adopting the line selection of YK2020 which covered 99 \% in the temperature range from 100~K to 1000~K, in addition to the CKD method.
It provided the maximum error of cooling rate less than 100 \% for CO but sometimes larger than about 100 \% for H$_2$O even at the atmospheric temperature less than 1000~K (see Fig.~\ref{fig:atom_comp2}), in which temperature range weaker lines have a smaller contribution to the cooling rate. 
Although we selected strong lines covering 99 \% of the total emission energy in 
the temperature range, the selection was done under an optically thin condition.
Thus, it cannot be accurate where the optical thickness of the strong lines becomes large and then the sum of the other neglected lines becomes dominant, as shown in Fig.\ref{fig:atom_comp_d1}$d$, especially at radial distances smaller than 2~$R_p$. 

It is necessary to increase the number of the selected lines in order to improve the accuracy of cooling rate estimation and to extend the considered temperature range in the LS method. 
If we select strong lines covering 99.9 \% of the total emission energy at $T=$100-1000~K, the number of lines increases from 638 to 1140 for CO which improves the maximum error of the cooling rate in the LS method to 33 \% from 51 \% for the H$_2$-CO isothermal atmosphere with $T=$1000~K.
Likewise, in the case of H$_2$O, the number of lines increases from 5021 to 22230  which improves the maximum error to 25 \% from 88 \%.  
Since the numerical cost is proportional to the number of lines, the numerical cost increases by a factor of 1.8 for CO and 4.4 for H$_2$O.
Additionally, to extend the temperature range to 100-2000 K, even more lines are necessary, leading to a selection of 930 lines for CO and 21030 for H$_2$O with covering 99~\% of the total emission energy at this extended range.

Because of the trade-off in the LS method, the LS method works better for species with fewer lines. 
Such a trend was obtained in the comparison between CO and H$_2$O at Sec.~\ref{sec:res}. 
In Fig~\ref{fig:atom_comp2} and Tab.~\ref{tab:cost}, the LS method provided more accurate cooling rates with a lower numerical cost for CO than for H$_2$O which has about 100 times more lines (half a million) than CO ($\sim$7000) in {{\it{HITRAN}}} database \citep{GORDON+2022}. 
As examples of the other molecules in the {{\it{HITRAN}}} database \citep{GORDON+2022}, N$_2$ and CS have thousands lines like CO while CO$_2$ and CH$_4$ have about half a million lines like H$_2$O. 
Monoatomic molecules don't have rotational-/vibrational-transitions and then have a much fewer lines than polyatomic molecules. 
Thus, the LS method, often employed in hydrodynamic simulations and upper atmospheric models to consider atomic line cooling \citep[e.g.,][]{Murray-Clay+2009,Salz+2016, Nakayama+2022}, would be an effective approach for these  molecules but not for molecules with many lines.

\subsection{Applications to other condition}
\label{sec:disc4}
The CKD method may be also useful in hydrodynamic simulations for astronomical environments other than atmospheres.
For instance, a primordial star forming cloud can be optically thick in H$_2$ line emission \citep[e.g.,][]{Oumukai+1998,Ripamonti+2002}.
\citet{Greif2014} showed the importance of smaller cross sections in H$_2$ line wings for cooling rate calculations in the hydrodynamic simulations of optically thick star forming clouds.
For such a case, the CKD method may be useful to consider the H$_2$ line wings and the overlapping between themselves.

\section{Summary and Conclusion}
\label{sec:conc}
In this study, we have investigated how efficiently the CKD method works and the importance of considering all lines of molecules for H$_2$-dominant
transonic atmospheres containing H$_2$O or CO with a one-dimensional atmospheric model.
Our radiative transfer simulations have demonstrated that the CKD method with R~$\geq$~1000 mostly agrees with the LBL method in cooling rate and the sum of the weak lines of H$_2$O can become the primary source of radiative cooling in optically thick regions.
Also, the simulations have shown that the CKD method works better for the H$_2$O-enriched atmospheres than for the CO-enriched atmospheres because H$_2$O has much more lines than CO. 
Additionally, our hydrodynamic simulations adopting the CKD method with R of 100--3000 have shown that the difference in the escape rate for each R is within 50~\%.
Also, we found the CKD method with R of 1000 is effective for the escape rate and temperature profile calculation, as the difference between those with R$=$1000 and  R$=$3000 is smaller than 10~\%.

As demonstrated in Sec.~\ref{sec:res}, the CKD method can be applied to upper atmospheres and is a powerful tool for hydrodynamic models to consider many lines of molecules without any prior selections for effective lines.
As it has been widely used for simulations of lower regions of various atmospheres such as CO$_2$-enriched, H$_2$O-enriched
and N$_2$-O$_2$-enriched
atmospheres, it can be used for the simulations of upper atmospheres for various exoplanets such as hot-Jupiter and super-Earths. 
This method is especially useful for simulating heavy-element-enriched atmospheres, since hydrodynamic simulations need to consider the radiation of many lines of molecules as coolants for them.
In addition, this method is more useful in considering a wide temperature range than the LS method. This is because the numerical cost of the CKD method linearly increases with the increase of its temperature grid points, but that of the LS method linearly increases with the number of lines and thus significantly increases with a temperature range (see Sec.~\ref{sec:disc3}).
Note that, our investigation is under the assumption of a LTE condition. As discussed in Sec.~\ref{sec:disc21},  the lack of collisional excitation rates in the current database makes it difficult to consider non-LTE effects for all lines and then hard to get an advantage of the CKD method.

%\begin{acknowledgments}
\acknowledgments
We appreciate Gen Chiaki for giving fruitful discussion about cooling rate calculations and possible applications of this study. We also thank Ryohei Nakatani for helpful discussion.
This work was supported by JSPS KAKENHI grant 22K14090 and 23K13161. TY was also supported by JSPS KAKENHI grant No. 23KJ0093.
%\end{acknowledgments}

\appendix
\section{Appendix information}
\begin{figure*}
 \begin{minipage}{0.5\textwidth}
    \begin{center}
        (a) CO
  \includegraphics[width=\textwidth]{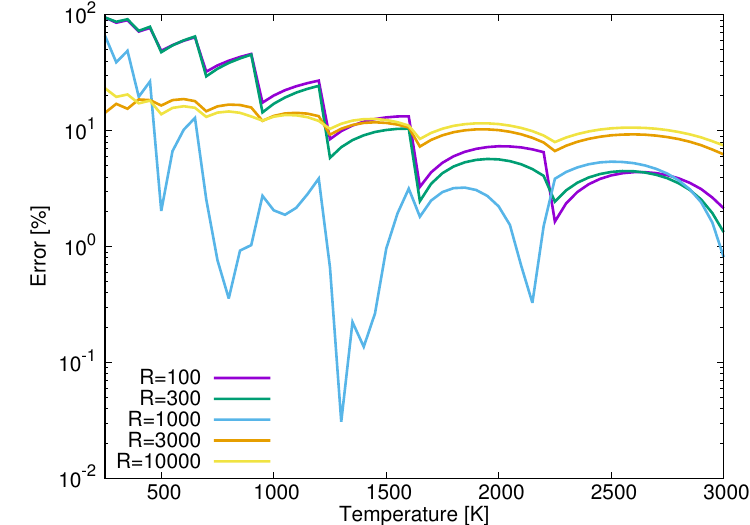}
   \end{center}
 \end{minipage}
   \begin{minipage}{0.5\textwidth}
    \begin{center}
        (b) \rm{H$_2$O}
  \includegraphics[width=\textwidth]{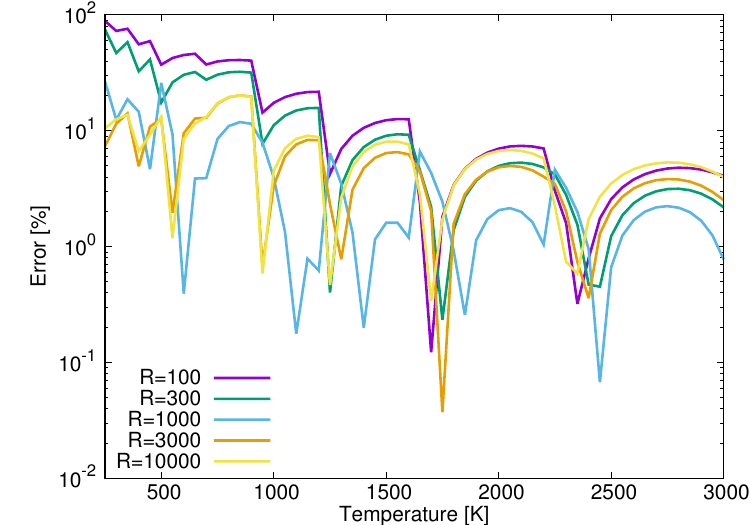}
   \end{center}
 \end{minipage}
\caption{Error in cooling rate calculations with CKD method for CO and \rm{H$_2$}O in an optically thin condition are shown as functions of temperature. For the resolution of the CKD method, five values are chosen: 100 (purple), 300 (green), 1000 (light blue), 3000 (orange) and 10000 (yellow).
 Right-side and left-side panels respectively show the error for CO and $\rm{H_2}$O. 
}
\label{fig:atom_comp3}
\end{figure*}

In Fig~\ref{fig:atom_comp2}, we showed the maximum error, $\epsilon_{max}$, of the CKD method increases with R for R $\geq1000$ in the cases for CO with $T=2000$~K and H$_2$O with $T\geq500$~K. In these cases, we found that the maximum error appeared at the larger distances where were in optically thin conditions in our simulation.
In order to clarify the error dependence of the CKD method on R at optically thin conditions, we directly calculate the cooling rate at such conditions. Fig.~\ref{fig:atom_comp3} shows the error in cooling rates of CO ($a$) and H$_2$O ($b$)  calculated by the CKD method with R$=100$--10000.
As shown in ~\ref{fig:atom_comp3}$a$ and $b$, R$=3000$ or 10000 give the smallest error in  temperature less than $\sim500$~K. On the other hand, in a temperature range from $\sim500$~K to 2000~K, R$=1000$ gives the smallest error and R$=3000$ or 10000 becomes less accurate than it. As mentioned in Sec.~\ref{sec:res}, this is because larger R reduces the number of lines in each wavelength bin and thus, larger R inaccurately
estimates the averaged intensity due to the dispersion of line intensity. Note that, there are many dips in the error over temperature, that present at the same temperature with the temperature grid of cross-section data. 

\begin{figure*}
 \begin{minipage}{0.5\textwidth}
      \begin{center}
 $\rm{H_2}$+CO atmospheres
 \end{center}
    \begin{center}  
    \includegraphics[width=\textwidth]{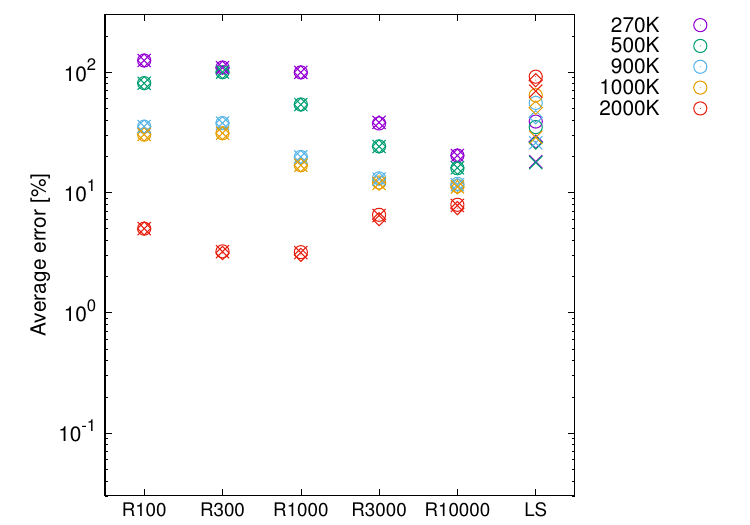}
   \end{center}
 \end{minipage}
   \begin{minipage}{0.5\textwidth}
          \begin{center}
$\rm{H_2}$+$\rm{H_2}$O atmospheres
 \end{center}
    \begin{center}
    \includegraphics[width=\textwidth]{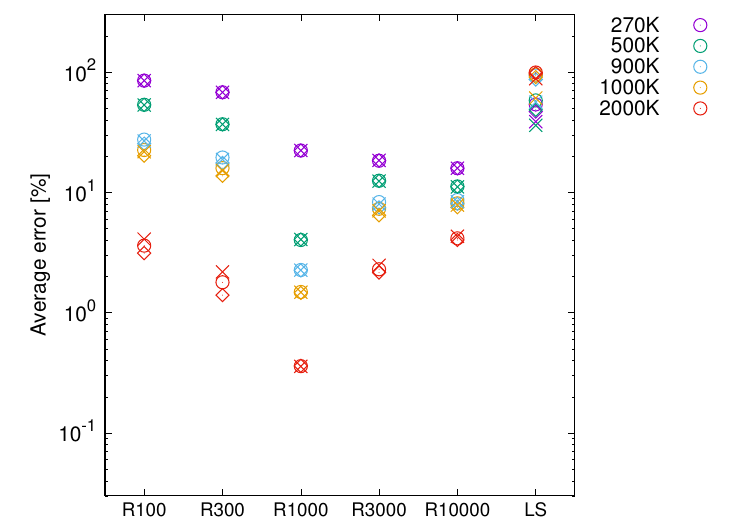}
   \end{center}
 \end{minipage}
 \begin{minipage}{0.5\textwidth}
    \begin{center}
  \includegraphics[width=\textwidth]{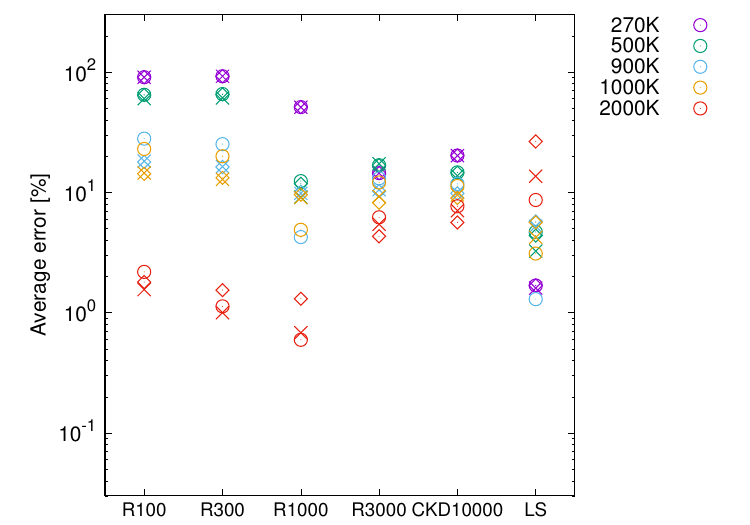}
   \end{center}
 \end{minipage}
  \begin{minipage}{0.5\textwidth}
    \begin{center}
  \includegraphics[width=\textwidth]{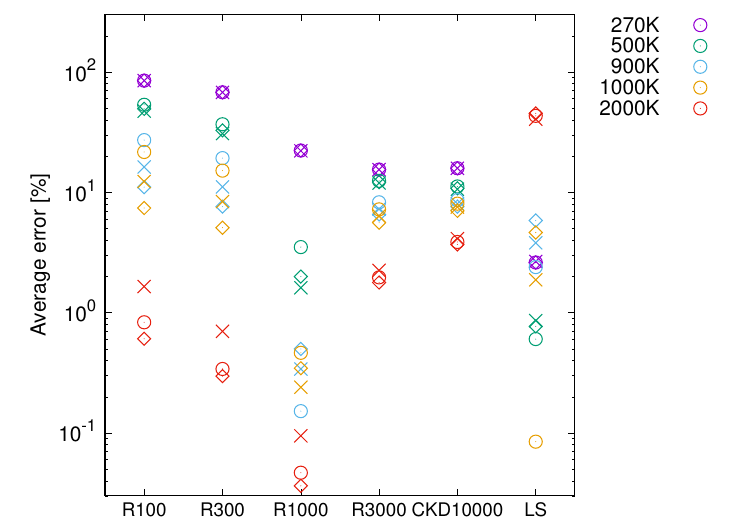}
   \end{center}
 \end{minipage}
\caption{Error in cooling rate calculations for CO and \rm{H$_2$}O in \rm{H$_2$}-dominant
atmospheres with temperatures of 270~K (purple), 500~K (green), 900~K (light blue), 1000~K (orange) and 2000~K (red) are shown.  Correlated-K method for five choices of resolution of 100, 300, 1000, 3000 and 10000, and the LS method are used for the cooling rate calculations, which are shown in  horizontal axis.
Upper and lower panels respectively show the maximum error and the space-averaged error, while left-side and right-side panels respectively show the two error for \rm{H$_2$}-dominant
atmospheres containing only CO and only $\rm{H_2}$O. 
In the atmospheres, the ratios of each two species (i.e., CO and \rm{H$_2$}O) and \rm{H$_2$}
are assumed to be 1 (circle), 0.1 (diamond) and 0.01 (cross).
}
\label{fig:atom_comp4}
\end{figure*}

Fig.\ref{fig:atom_comp4} shows $\epsilon_{max}$ and $\epsilon_{ave}$ in the cooling rate calculations for CO and \rm{H$_2$}O in \rm{H$_2$}-dominated atmospheres with the radiative species fraction of 1, 0.1 and 0.01, while Fig~\ref{fig:atom_comp2} showed only those for only the radiative species fraction of 0.1. As mentioned in Sec.~\ref{sec:res}, $\epsilon_{max}$ is almost the same as that for the fraction of 0.1 while $\epsilon_{ave}$ scatters a little bit. 

\begin{figure*}
\begin{center}
\includegraphics[width=\columnwidth]{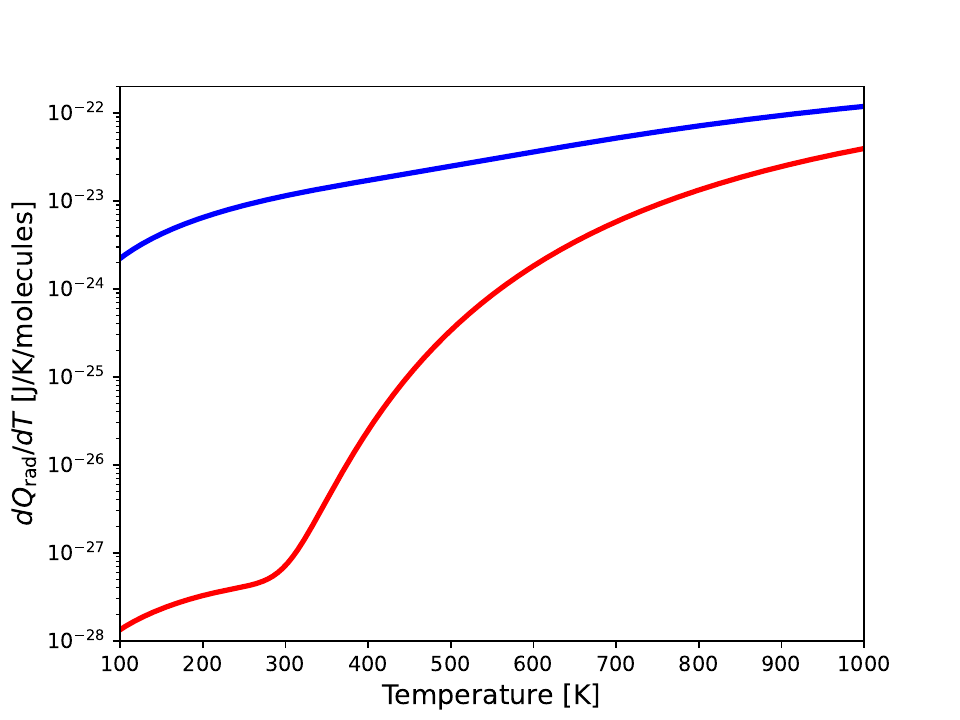}
\caption{Differential coefficient of the radiative cooling rate of H$_{2}$O (blue) and CO (red) per one molecule in optically thin condition as a function of temperature.}
\label{fig:dcool_comp}
\end{center}
\end{figure*}

Fig.\ref{fig:dcool_comp} shows the temperature derivative value of $Q_{\rm rad}$ per one CO and H$_{2}$O molecule at optically thin condition. As shown in Fig.\ref{fig:dcool_comp}, since that of H$_{2}$O is much larger than CO, the cooling rate of H$_{2}$O is much more sensitive to temperature than that of CO.

\begin{figure*}
\begin{center}
 \begin{minipage}{0.45\textwidth}
    \begin{center}
    ($a$)
  \includegraphics[width=\textwidth]{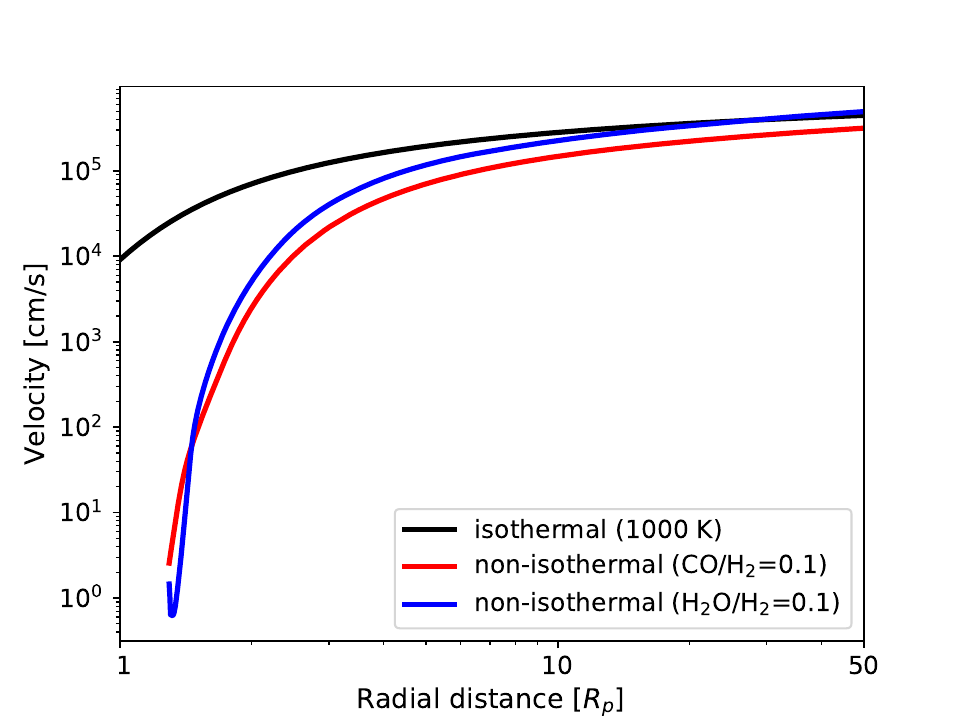}
   \end{center}
   \end{minipage}
   \begin{minipage}{0.45\textwidth}
    \begin{center}  
    ($b$)
    \includegraphics[width=\textwidth]{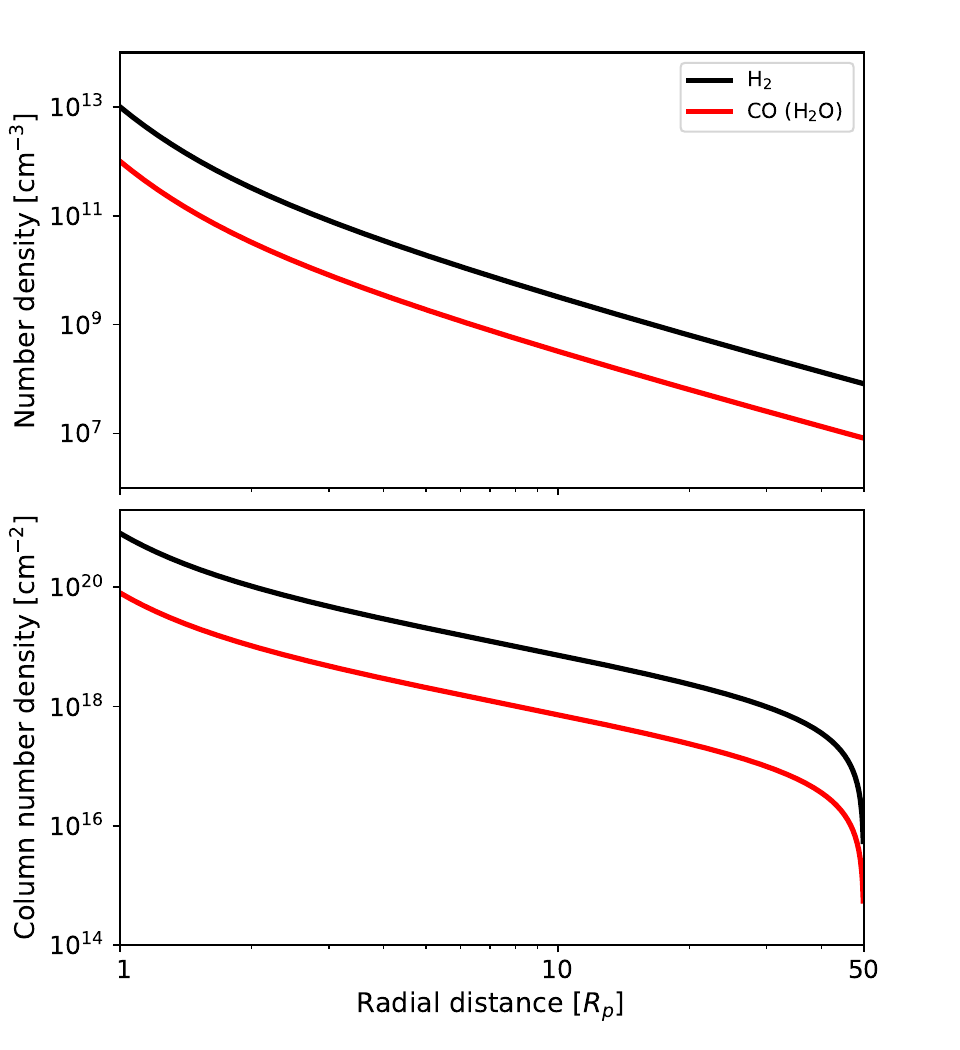}
   \end{center}
 \end{minipage}
 \\
 \begin{minipage}{0.45\textwidth}
    \begin{center}
    ($c$)
  \includegraphics[width=\textwidth]{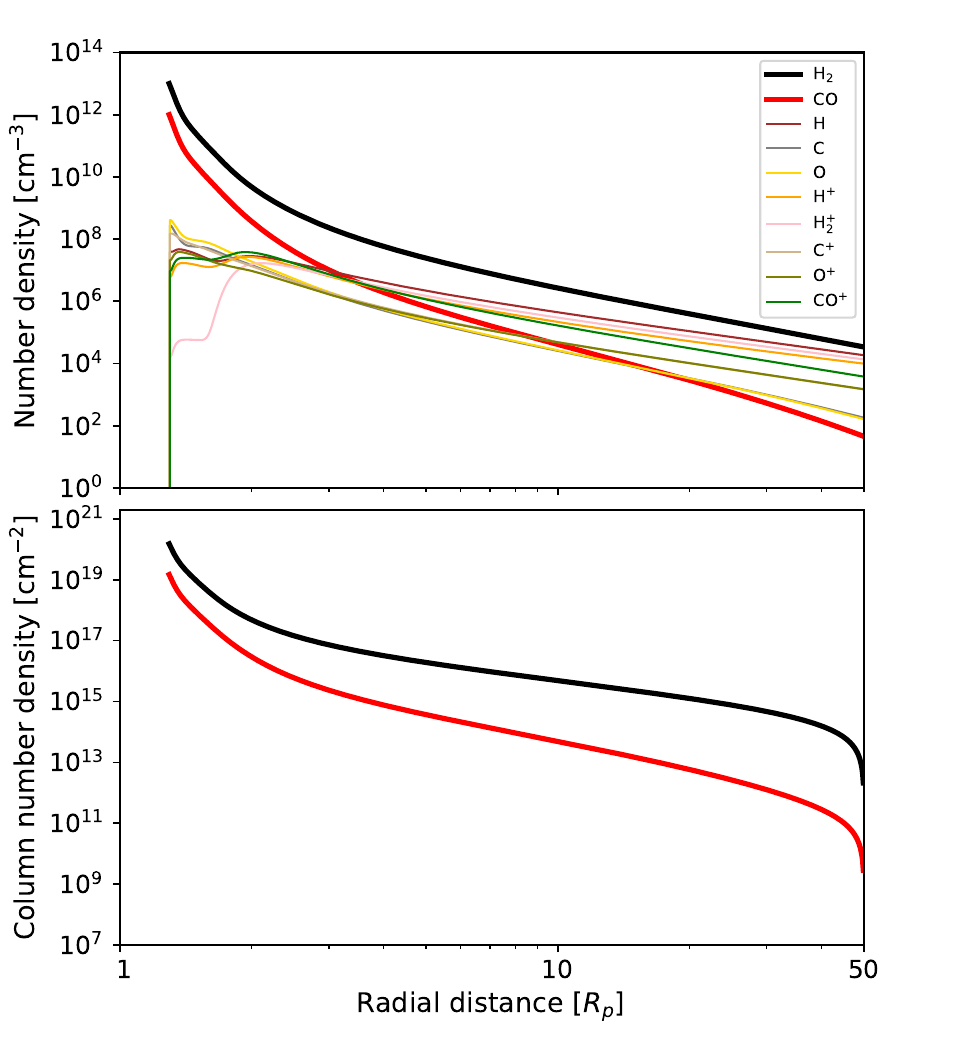}
   \end{center}
 \end{minipage}
  \begin{minipage}{0.45\textwidth}
    \begin{center}
    ($d$)
  \includegraphics[width=\textwidth]{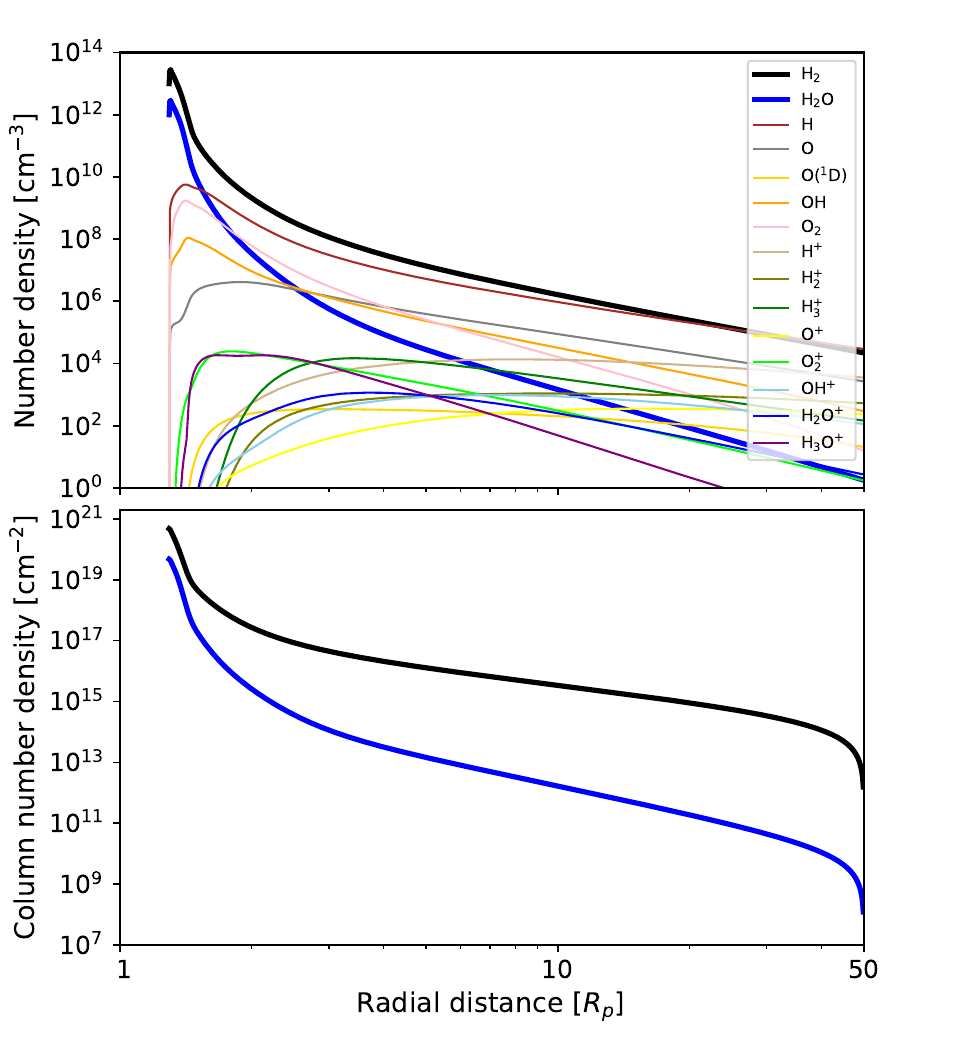}
   \end{center}
 \end{minipage}
\caption{(a) Mean velocity profiles of the isothermal atmosphere with $T=1000\,\mathrm{K}$ and CO(H$_{2}$O)/H$_{2}$=0.1 (black), the non-isothermal atmosphere with CO/H$_{2}$=0.1 (red), and the non-isothermal atmosphere with H$_{2}$O/H$_{2}$=0.1 (blue). (b)-(d) Number density and column number density profiles of the isothermal atmosphere with $T=1000\,\mathrm{K}$ and CO(H$_{2}$O)/H$_{2}$=0.1 (b), non-isothermal atmosphere with CO/H$_{2}$=0.1 (c), and non-isothermal atmosphere with H$_{2}$O/H$_{2}$=0.1 (d). Here the non-isothermal atmospheric profiles are obtained by applying the CKD method with a wavelength resolution of 1000. Note, for the isothermal atmosphere (b), the number and column densities of CO and H$_{2}$O are the same and thus their lines are overlapped.
}
\label{fig:num}
\end{center}
\end{figure*}

Fig.~\ref{fig:num} shows the mean velocity, number density and column number density profiles of the isothermal atmosphere with $T=1000\,\mathrm{K}$ and CO(H$_{2}$O)/H$_{2}$=0.1, and the non-isothermal atmospheres with CO(H$_{2}$O)/H$_{2}$=0.1 obtained by applying the CKD method with a wavelength resolution of 1000. 
As shown in Fig.~\ref{fig:num}$a$, the atmospheric velocity is radially accelerated to supersonic.
The composition profiles for the non-isothermal atmospheres show that H$_2$O and CO, key species for cooling rate, are gradually photo-ionized to produce various atoms and ions with altitudes(Fig.\ref{fig:num}$c$, $d$).
Compared with calculation results for Hot Jupiters’ atmospheres \citep[e.g.,][]{Garcia2007}, a significant fraction of molecules stays undissociated due to the lower UV irradiation and effective advection under low planetary gravity, as described in \cite{YK2020} and \cite{Yoshida+2022} (please see these studies for more details).
For the same reason, as we assumed the lower UV irradiation and lower gravity for the non-isohtermal H$_2$-H$_2$O atmosphere than those in \cite{Yoshida+2022}, the former made the photo-ionization/-dissociation rate lower and the latter made advection more effective. Then, as shown in Fig.~\ref{fig:num}$d$, the H$_2$O abundance of this simulation is larger than that shown in \cite{Yoshida+2022}. 
Also, as for the difference of the non-isohtermal H$_2$-CO atmosphere with \cite{YK2020}, we consider the presence of only CO and H$_2$ at the lower boundary condition but \cite{YK2020} considered CH$_4$ as well. Compared to the results for CO/H$_2$=CH$_4$/H$_2$=0.1 in \cite{YK2020}, CO remains more undissociated in our simulation (Fig.~\ref{fig:num}$c$) because of more effective advection caused by the lower atmospheric mean molecular weight and no radiative cooling from CH$_4$.

\bibliography{ref}{}
\bibliographystyle{aasjournal}

\end{document}